\documentclass{SciPost}
\usepackage{etoolbox}
\usepackage{mathtools}
\usepackage{dcolumn}
\usepackage{bm}
\usepackage{verbatim}
\usepackage[english]{babel}
\usepackage{graphicx}%
\usepackage{multirow}%
\usepackage{amsmath}%
\usepackage[title]{appendix}%
\usepackage{xcolor}%
\usepackage{textcomp}%
\usepackage{booktabs}%
\usepackage{listings}%
\usepackage{bbold}
\usepackage{amsthm}
\usepackage{booktabs}
\usepackage{array}
\usepackage{makecell}
\usepackage{multirow}
\usepackage{siunitx}
\usepackage[capitalise]{cleveref}

\DeclareMathOperator{\am}{am}
\DeclareMathOperator{\sn}{sn}
\DeclareMathOperator{\cd}{cd}
\DeclareMathOperator{\nc}{nc}
\DeclareMathOperator{\cs}{cs}

\DeclareMathOperator{\sd}{sd}
\DeclareMathOperator{\nd}{nd}
\DeclareMathOperator{\cn}{cn}
\DeclareMathOperator{\dn}{dn}
\DeclareMathOperator{\dc}{dc}
\DeclareMathOperator{\ds}{ds}
\DeclareMathOperator{\ns}{ns}

\hypersetup{
    colorlinks,
    linkcolor={red!50!black},
    citecolor={blue!50!black},
    urlcolor={blue!80!black}
}

\usepackage[bitstream-charter]{mathdesign}
\DeclareSymbolFont{usualmathcal}{OMS}{cmsy}{m}{n}
\DeclareSymbolFontAlphabet{\mathcal}{usualmathcal}

\fancypagestyle{SPstyle}{
\fancyhf{}
\lhead{\colorbox{scipostblue}{\bf \color{white} ~SciPost Physics }}
\rhead{{\bf \color{scipostdeepblue} ~Submission }}

\fancyfoot[C]{\textbf{\thepage}}
}

\definecolor{emerald}{rgb}{0.31, 0.78, 0.47}
\definecolor{blue(ncs)}{rgb}{0.0, 0.53, 0.74}

\begin{document}

\pagestyle{SPstyle}

\begin{center}{\Large \textbf{\color{scipostdeepblue}{
The Yang-Baxter Equation for the Chiral Potts Model and Integrable Parafermions\\
}}}\end{center}

\begin{center}\textbf{
Zhao Zhang\textsuperscript{1$\star$} 
}\end{center}

\begin{center}

{\bf 1} Department of Physics, University of Oslo, P.O. Box 1048 Blindern, N-0316 Oslo, Norway
\\[\baselineskip]
$\star$ \href{mailto:email1}{\small zhao.zhang@fys.uio.no}\,
\end{center}

\section*{\color{scipostdeepblue}{Abstract}}
\textbf{\boldmath{A new type of Yang-Baxter equation (YBE) for $R$-operators depending on three spectral parameters is constructed from the star-triangle relation for the chiral Potts model. As the $Z_N$ symmetric generalization to the Ising model, its Boltzmann weights are known to depend on two variables describing a curve with genus larger than one for $N>2$, except for the self-dual point corresponding to the Fateev-Zamolodchikov chain. Combined with the fact that quantum Hamiltonians of edge-type models such as the Ising model contain both nearest-neighbor interaction and onsite potential terms, this leads naturally to an additional spectral parameter in the associated $R$-operator. The construction extends the edge-vertex correspondence of solvable lattice models, and provides a bridge between the Bazhanov-Stroganov four-parameter $R$-matrix---realized as an intertwiner of cyclic representations of $U_q(\mathfrak{sl}_2)$ at a root of unity---and Shastry's two-parameter $R$-operator obtained from the decorated YBE.
}}

\vspace{\baselineskip}

\noindent\textcolor{white!90!black}{%
\fbox{\parbox{0.975\linewidth}{%
\textcolor{white!40!black}{\begin{tabular}{lr}%
  \begin{minipage}{0.6\textwidth}%
    {\small Copyright attribution to authors. \newline
    This work is a submission to SciPost Physics. \newline
    License information to appear upon publication. \newline
    Publication information to appear upon publication.}
  \end{minipage} & \begin{minipage}{0.4\textwidth}
    {\small Received Date \newline Accepted Date \newline Published Date}%
  \end{minipage}
\end{tabular}}
}}
}


\vspace{10pt}
\noindent\rule{\textwidth}{1pt}
\tableofcontents
\noindent\rule{\textwidth}{1pt}
\vspace{10pt}

\section{Introduction}
\label{sec:intro}

The laws of physics are typically expressed as differential equations that govern instantaneous evolution. In order to be of any predictive power, they must be solved or integrated. This is straightforward for a limited class of few-body systems (except when chaos intervenes), but rarely feasible in statistical mechanics or quantum many-body physics. One therefore often relies on perturbation theory, which is ill-suited to uncover novel and exotic phenomena in strongly interacting systems. In this context, integrability and exact solvability play a central role in revealing new physics. However, its success seems to have been overshadowed by an Anna Karenina principle of integrability: All perturbation theories are alike, each integrable model is integrable in its own way.

In statistical mechanics, lattice models are commonly classified into three types—edge, vertex, and face models—according to where the Boltzmann weights (BWs) are assigned \cite{baxter2013exactly}. A canonical example of an edge model is the Ising model, first solved by Onsager in 1944 \cite{PhysRev.65.117}. Although only briefly remarked in his original solution, the star-triangle transformation is really at the core of his elliptic function parametrization \cite{RevModPhys.17.50,mills_ascher_jaffee_1971}. This transformation implies the commutation of transfer matrices which was employed in his unpublished derivation of the spontaneous magnetization that was later calculated by Yang using a different method \cite{PhysRev.85.808}. A version of the same transformation commonly appears in the analysis of resistor networks, where it was introduced by Kennelly in 1899 \cite{kennelly99}. In fact, that transformation is also related to the Potts and Gaussian models in statistical mechanics \cite{RevModPhys.54.235,perk2006yang}.

Unlike the diagonal-to-diagonal transfer matrix of edge models defined as a product of alternating horizontal and vertical bonds on a square lattice, the transfer matrix of vertex models are defined as a product of vertex BWs along a single row. The row-to-row transfer matrices mutually commute when the $R$-matrices specifying the vertex BWs satisfy the Yang-Baxter equation (YBE). An $R$-operator can be considered an integral form of the local Hamiltonian, as it is the building block of an integrable quantum circuit that describes the time evolution of the one-dimensional (1D) quantum many-body system \cite{Miao2023integrablequantum}. The YBE was first discovered by Yang in solving the 1D fermions with repulsive delta-function interaction \cite{PhysRevLett.19.1312} as the consistency equation for factorizable scattering $S$-matrices. A similar equation in real space is satisfied by the $R$-matrix of Baxter's eight-vertex model \cite{PhysRevLett.26.832} and a procedure is summarized for finding such solvable models by inserting spectral parameters to braid group representations obeying the Temperley-Lieb algebra \cite{doi:10.1142/S021797929000036X}.

While the integrability of vertex and face models are both underpinned by the YBE, Onsager's star-triangle relation (STR) responsible for the solvability of edge models is manifestly different, even though the terminology has always been used interchangeably in the literature. Despite numerous efforts in mapping one to the other over the years \cite{pokrovskyBashilov,Miao2023integrablequantum,MARTINS2024116610,BAZHANOV2023116055}, it was not until the year the world lost both Yang and Baxter that Martins made the major progress towards elegantly unifying the two cornerstones of their separately mansions \cite{MartinsRmat}. The moral of the story of his seminal work is that the $R$-operator for edge models has one more spectral parameter in its argument than the BWs, as vertical and horizontal bonds are treated anisotropically in the correspondence between 2D classical to 1D quantum models.  

There are in fact two distinct reasons why an $R$-operator or BW may depend separately on more than one spectral parameter. The number of spectral parameters has important consequence on the complexity of the problem, as relativistic $R$-matrices that depend only on the difference between two spectral parameters can be iteratively bootstrapped from the Hamiltonian \cite{10.21468/SciPostPhys.20.4.102} satisfying the Reshetikhin condition \cite{Kulish:1982aa}. Two well understood instances of non-relativistic integrable Hamiltonians with bivariate $R$-operators are the Hubbard model and the XYh chain \cite{PhysRevLett.56.2453,ShastryDYBE, FermiR,RmatforHubbard, zhang2026integrablefreeinteractingfermions}. By now, it is fairly obvious that the extra spectral parameter can be attributed to the onsite potentials in these Hamiltonians that are absent in integrable spin chains corresponding to vertex models like the XYZ model. Meanwhile, among edge models where an onsite potential always emerge in the classical-quantum correspondence, it has been long known that the BWs themselves can depend on two spectral parameters, and an archetypical example is provided by the chiral Potts model (CPM) \cite{AuMcCoyPerkTangYan,McCoyPerkTangSah,AuMcCoyPerkTang,BaxterPerkAu,masterkey}.

In contrast to the more pedestrian generalization to the Ising model, the $S_N$-symmetric Potts model, the CPM has the smaller $Z_N$ symmetry and is often referred to as the clock model, as the quantum Hamiltonian are expressed in terms of clock and shift operators generalizing the Pauli $Z$ and $X$ matrices. Unlike the Ising case, the spectral parameters of the CPM lie on higher genus curves, which prevents a single-valued global rapidity parameterization. The importance of the CPM is also further underscored by its close connection to $Z_N$ parafermions \cite{Fateev:1985mm,FZ,HowesKadanoffNijs,GehlenRittenberg,Fendley_2012,PhysRevB.94.165142}. When the CPM with bivariate BWs is incorporated into Martins' embedding for edge models into vertex ones, the two distinct sources of non-relativity for the $R$-operator coexists, yielding one that depends on three spectral parameters. This case was not treated in Martins' original work \cite{MartinsRmat}, although he did examine the $Z_N$ Fateev-Zamodlochikov (FZ) model \cite{FZ} and various higher-spin generalizations of the Ising model, all with BWs depending on a single parameter.

To fully appreciate Martins' construction and its extension to the CPM and integrable parafermions, it is important to distinguish between an $R$-matrix and an $R$-operator. The former satisfies algebraic YBEs as matrix identities, while the latter obeys operator YBEs acting on three adjacent sites of a quantum chain. Although an $R$-matrix defines a 2D statistical mechanical model whose transfer matrix generates the total Hamiltonian of the corresponding quantum chain via differentiation, the $R$-matrix itself is not related to the local Hamiltonian density. Thus, in the context of integrable quantum circuits for 1D quantum systems, only $R$-operators constructed via Martins' or Shastry's approaches can be interpreted as unitary quantum gate corresponding to integral forms of the local Hamiltonian operators. 

Using this terminology, the two-BW product of Ref.~\cite{pokrovskyBashilov} and four-BW product of Ref.~\cite{BAZHANOV2023116055}---with one and four spectral parameters respectively---both yield $R$-matrices for the Ising model that satisfy algebraic YBEs, but without a direct link to local quantum Hamiltonian operators. Ref.~\cite{Miao2023integrablequantum} does construct three-parameter $R$-operators for the Ising Hamiltonian that satisfies an operator YBE. However, for same Hamiltonian a two-parameter $R$-operator can be obtained either from Shastry's method or equivalently from Martins' construction. In this sense, the three-parameter formulation of Ref.~\cite{Miao2023integrablequantum} appears nonminimal. 

\begin{figure}[t]
    \centering
    \includegraphics[width=\linewidth]{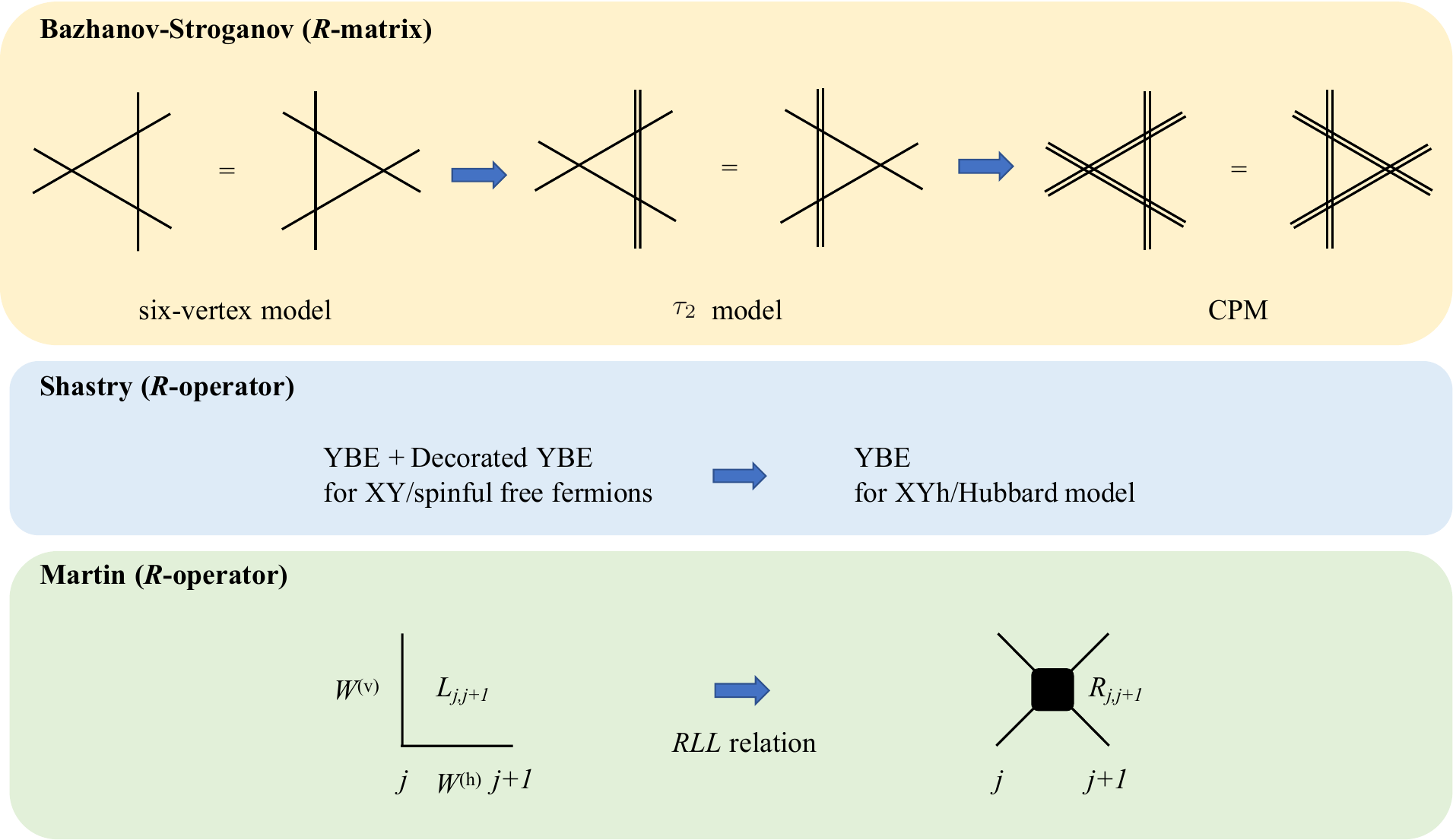}
    \caption{Comparison of the three different approaches for finding non-relativistic $R$-matrices or $R$-operators.}
    \label{fig:comparison}
\end{figure}

Turning to the CPM, Bazhanov and Stroganov constructed a four-parameter $R$-matrix from the intertwiner of cyclic representations of the quantum group $U_q(\mathfrak{sl}_2)$ at a root of unity, which can be expressed as a product of four BWs \cite{BazhanovaStroganov, JimboMiwa}. An intermediate step defines the $\tau_2$ or Baxter-Bazhanov-Strognanov (BBS) model, which provides a bridge between the six-vertex model and the CPM. Although their $R$-matrix can be shown to satisfy an algebraic YBE, it does not directly define an $R$-operator satisfying a operator YBE. Nevertheless, I will show that upon identifying two of the four spectral parameters, it reduces to the matrix elements of the $R$-operator obtained from extending Martins' construction. Since the extended $R$-operator also satisfies a generalized form of Shastry's decorated YBE \cite{ShastryDYBE,FermiR,zhang2026integrablefreeinteractingfermions} for parafermions, the resulting three-parameter $R$-operator simultaneously unifies two earlier constructions and generalizes Martins' more recent one. The three different constructions are summarized in Fig.~\ref{fig:comparison}.

The remainder of the paper is organized as follows. In Sec.~\ref{sec:IM}, I review Shastry's construction specialized to the transverse-field Ising model (TFIM), viewed as a special case of the XYh chain, and demonstrates how the resulting $R$-operator, expressed in terms of the BWs, matches the result of Martins' edge-vertex conversion. Sec.~\ref{sec:ZNparafermion} reviews the $Z_N$ generalization to the TFIM, namely the superintegrable Hamiltonian of von Gehlen and Rittenberg \cite{GehlenRittenberg}, which remains integrable for arbitrary external field strength. Its classical counterpart belongs to the class of CPMs, which are summarized in Sec.~\ref{sec:CPM} with emphasis on the superintegrable case. The main result is presented in Sec.~\ref{sec:STLYBE}, where Martins' method is extended to the CPM, yielding a new type of operator YBE for $R$-operators depending on three spectral parameters. In the standard basis, this operator is shown in Sec.~\ref{sec:tau2} to be related to the four-parameter $R$-matrix of Bazhanov-Stroganov by identifying two of its parameters. In the Schwinger basis, it is connected in Sec.~\ref{sec:DYBE} to generalizations of Shastry's decorated YBE. Finally,  Sec.~\ref{sec:conclusion} offers conclusions and an outlook on possible extensions to parafermionic versions of the XYh and Hubbard chains, pointing in particular to Fock parafermion operators introduced by Cobanera and Ortiz \cite{PhysRevA.89.012328} and reviewed in detail in Appendix \ref{sec:Fock}.

\section{The Ising model and Majorana Fermions} \label{sec:IM}

In Ref.~\cite{zhang2026integrablefreeinteractingfermions}, an $R$-operator for the XYh model was obtained following the earlier work of Ref.~\cite{RmatforHubbard}. The special case when the coupling in the $y$-direction vanishes reduces to the Ising model after a duality transformation that interchanges order and disorder operators. Here we repeat that derivation in the conventional basis for the Ising Hamiltonian
\begin{equation}
    H_\mathrm{IM}=-\sum_j Z_jZ_{j+1}-U\sum_j X_j. \label{eq:Ising}
\end{equation}The Majorana fermions are introduced by the Jordan-Wigner (JW) transformation
\begin{equation}
    a_j=\left(\prod_{k=1}^{j-1}X_k\right)Z_j,\quad b_j=-ia_jX_j=-i\left(\prod_{k=1}^{j-1}X_k\right)Z_jX_j,
\end{equation}which satisfy the anticommutation relations
\begin{equation}
    \{a_j,a_k\}=\{b_j,b_k\}=2\delta_{jk},\quad \{a_j,b_k\}=0, \quad \forall\ j,k.
\end{equation}In terms of the Majorana fermions, \eqref{eq:Ising} is rewritten as
\begin{equation}
    H_\mathrm{IM}=-i\sum_j b_ja_{j+1}-iU\sum_j a_jb_j. \label{eq:Majorana}
\end{equation}

\subsection{Constructing interacting \texorpdfstring{$R$}{R}-matrix from decorated Yang-Baxter equations}\label{sec:review}

If $U=0$, local degrees of freedom decouple, and the non-interacting $R$-operator is given by \cite{YuGe1}
\begin{equation}
    \check{R}_{j,j+1}(\mu)=e^{-i\mu b_ja_{j+1}}=\cosh\mu-ib_ja_{j+1}\sinh\mu , \label{eq:freeR}
\end{equation}which satisfies the regularity $\check{R}_{j,j+1}(0)=1$, inversion relation $\check{R}_{j,j+1}(\mu)\check{R}_{j,j+1}(-\mu)=1$, and reproduces the Hamiltonian by $\partial_\mu \check{R}_{j,j+1}(\mu)|_{\mu=0}=-ib_ja_{j+1}$. The operator $X_j=-ia_jb_j$ has the property $X_j\check{R}_{j,j+1}(\mu)X_j=X_{j+1}\check{R}_{j,j+1}(\mu)X_{j+1}=\check{R}_{j,j+1}(-\mu)$, and
\begin{equation}
    \begin{split}
        [\check{R}_{j,j+1}(\mu),X_j+X_{j+1}]=&2i\sinh\mu (a_ja_{j+1}-b_jb_{j+1}),\\
        \{\check{R}_{j,j+1}(\mu),X_j-X_{j+1}\}=&2\cosh\mu (a_jb_{j}-a_{j+1}b_{j+1}).
    \end{split}
\end{equation}So according to the general framework detailed in Ref.~\cite{zhang2026integrablefreeinteractingfermions}, the $R$-operator for nonvanishing $U$ is given by 
\begin{equation}
    \check{R}_{j,j+1}(\mu,\nu)=\check{R}_{j,j+1}(\mu-\nu)+f(\mu,\nu)\check{R}_{j,j+1}(\mu+\nu)X_j,\label{eq:IsingR}
\end{equation}where $f(\mu,\nu)$ satisfies
\begin{equation}
  \frac{f(\mu,0)}{1-f(\mu,0)^2}\sinh(2\nu)=\frac{f(\nu,0)}{1-f(\nu,0)^2}\sinh(2\mu),
\end{equation}and
\begin{equation}
    f(\mu,\nu)=\frac{\cosh(\mu-\nu)}{\cosh(\mu+\nu)}\frac{f(\mu,0)-f(\nu,0)}{1-f(\mu,0)f(\nu,0)}.\label{eq:f}
\end{equation}These equations are uniquely solved by 
\begin{equation}
    f(\mu,\nu)= \frac{\sqrt{1+U^2\sinh^2(2\mu)}-\sqrt{1+U^2\sinh^2(2\nu)}}{U\sinh(2(\mu+\nu))}
\end{equation}such that it satisfies the initial condition $\partial_\mu f(\mu,\nu)|_{\mu=\nu}=U$.\footnote{The apparent difference from the expression in Ref.~\cite{zhang2026integrablefreeinteractingfermions} is due to a different normalization of $\check{R}_{j,j+1}(\mu)$.} Since \eqref{eq:freeR} has the composition property $\check{R}_{j,j+1}(\mu)\check{R}_{j,j+1}(\nu)=\check{R}_{j,j+1}(\nu)\check{R}_{j,j+1}(\mu)=\check{R}_{j,j+1}(\mu+\nu)$, \eqref{eq:IsingR} can be written as 
\begin{equation}
    \check{R}_{j,j+1}(\mu,\nu)=\check{R}_{j,j+1}(\mu)(1-ia_jb_jf(\mu,\nu))\check{R}_{j,j+1}(-\nu), \label{eq:IsingR2}
\end{equation}which could be renormalized by a factor of $(1-f(\mu,\nu)^2)^{-\frac{1}{2}}$ to satisfy the usual normalization $\check{R}_{j,j+1}(\mu,\nu)\check{R}_{j,j+1}(\nu,\mu)=1$.

At the critical point $U=1$, $f(\mu,\nu)= \frac{\sinh(\mu-\nu)}{\cosh(\mu+\nu)}$, and the $R$-operator simplifies to
\begin{equation}
    \check{R}_{j,j+1}(\mu,\nu)=\cosh(\mu-\nu)-i\sinh(\mu-\nu)(a_jb_j+b_ja_{j+1})+\sinh(\mu-\nu)\tanh(\mu+\nu)a_ja_{j+1}.
\end{equation}Notice that this interacting $R$-operator has not been derived previously \cite{KorepinIsing}. Its derivative with respect to the spectral parameter gives rise to a non-Hermitian integrable Hamiltonian
\begin{equation}
\begin{split}
    h_{j,j+1}(\nu)=&\partial_\mu \check{R}_{j,j+1}(\mu,\nu)|_{\mu=\nu}\\
        =&-i(a_jb_j+b_ja_{j+1})+\tanh(2\nu)a_ja_{j+1}\\
            =&-Z_jZ_{j+1}-X_j-\tanh(2\nu)X_jZ_jZ_{j+1}.
\end{split}
\end{equation}Similar non-Hermitian integrable deformation to the Ising model exists for general $U$ \cite{zhang2026integrablefreeinteractingfermions}.

\subsection{Expression in terms of Boltzmann weights}

To connect the $R$-operator obtained above from Shastry's method to the result of Martins' construction \cite{MartinsRmat}, we express it in terms of Onsager's elliptic function parametrization of the Ising model \cite{baxter2013exactly}. Useful definitions and properties of the Jacobi theta and elliptic functions are reviewed in Appendix \ref{sec:elliptic}. Relating the spectral parameter to the inverse temperature $K=\beta$ and identifying the elliptic modulus with the transverse field strength 
\begin{equation}
    \kappa=\frac{1}{\sinh2K\sinh2\overline{K}}=\frac{H^2_1(0)}{\Theta^2_1(0)}=U,
\end{equation}the relative BWs between antiparallel and parallel neighboring spin pairs on horizontal and vertical bonds of the square lattice can be expressed respectively as
\begin{equation}
\begin{split}
    W(-K)=&e^{-2K}=\sqrt{1+\sinh^2 2K}-\sinh 2K=\cn (iu)+i\sn (iu)=\frac{\sqrt{\kappa'}H_1(iu)+iH(iu)}{\sqrt{\kappa}\Theta(iu)},\\
    \overline{W}(-K)=&e^{-2\overline{K}}=\frac{\sqrt{1+\kappa^2\sinh^2 2K}-1}{\kappa \sinh 2K}=\frac{1-\dn (iu)}{i\kappa\sn (iu)}=\frac{\Theta(iu)-\sqrt{\kappa'}\Theta_1(iu)}{i\sqrt{\kappa}H(iu)},
\end{split}\label{eq:ellipara}
\end{equation}where $\kappa'=\sqrt{1-\kappa^2}=(\Theta(0)/\Theta_1(0))^2$ is the conjugate modulus, and the parameter $u$ was introduced by $\sinh 2K=-i\sn (iu)$ or equivalently $2K=\am(u,\kappa')$ for comparison with \eqref{eq:CPMBWs} in Sec.~\ref{sec:CPM}. Using \eqref{eq:firstkind} in Appendix \ref{sec:elliptic}, we have
\begin{equation}
    u(K)=\int_0^{2K}\frac{d\theta}{\sqrt{1+\kappa^2\sinh^2\theta}},  \label{eq:ufunc}
\end{equation}which is real and satisfy $0<u<I'$ as long as $K, \overline{K}$ are real and positive. The RHSs of \eqref{eq:ellipara} can be extended to a bivariate parametrization
\begin{equation}
\begin{split}
     W(u,v)=&\frac{i\Theta_1(iu)H(iv)+\Theta(iu)H_1(iv)}{iH(iu)\Theta_1(iv)+H_1(iu)\Theta(iv)}=\frac{\cn(iv)+i\dn(iu)\sn(iv)}{\cn(iu)+i\sn(iu)\dn(iv)}\\
     =&\cn\big(i(u-v)\big)-i\sn\big(i(u-v)\big),\\
     \overline{W}(u,v)=&\frac{-\Theta(iu)\Theta_1(iv)+\Theta(iv)\Theta_1(iu)}{iH_1(iu)H(iv)+iH(iu)H_1(iv)}=\frac{\dn(iu)-\dn(iv)}{i\kappa\big(\sn(iu)\cn(iv)+\sn(iv)\cn(iu)\big)}\\
     =&i\kappa\sn\left(\frac{i(u-v)}{2}\right)\cd\left(\frac{i(u-v)}{2}\right),
\end{split}\label{eq:ellipara2}
\end{equation}such that \eqref{eq:ellipara} corresponds to $W(-K)=W(0,u(K)), \overline{W}(-K)=\overline{W}(0,u(K))$, since $H(0)=0$. Notice that the final expressions above shows that the BWs of the Ising model, as the $N=2$ exceptional case of the CPM discussed in Sec.~\ref{sec:CPM}, indeed can be parameterized by the difference of two spectral parameters.

If we define $\widehat{W}(K)=\frac{1-W(K)}{1+W(K)}=-\tanh K$, then $\overline{W}'(0)=-\kappa=-U=U\widehat{W}'(0)$, and \eqref{eq:freeR} can be written as $\check{R}_{j,j+1}(\mu)=\cosh(\mu)\left(1+\widehat{W}(\mu)Z_jZ_{j+1}\right)$. \eqref{eq:f} becomes $f(\mu,\nu)=\overline{W}\big(u(-\nu), u(-\mu)\big)$ with the function $u(\mu)$ defined in \eqref{eq:ufunc}. So the $R$-operator \eqref{eq:IsingR} obtained from Shastry's decorated YBE can be expressed in the standard basis of $2\times 2$ matrices in terms of the BWs as Martins has realized \cite{MartinsRmat}, just like the more general CPM model to be discussed in the following sections.

\section{The von Gehlen-Rittenberg \texorpdfstring{$\mathbb{Z}_N$}{ZN} parafermion chain}\label{sec:ZNparafermion}

For an $N$-state local Hilbert space $\mathbb{C}^N$, the Pauli $X$ and $Z$ matrices generalize to the `shift' and `clock' matrices originally introduced by Sylvester in 1882
\begin{equation}
    X=\begin{pmatrix}
        0 & 0 & \cdots &0 & 1 \\
        1 & 0 &\cdots& 0 & 0\\
        0 & 1 & \cdots &0 & 0\\
        \vdots & \vdots & \ddots& \vdots & \vdots\\
        0 & 0 & \cdots &1 & 0
    \end{pmatrix}, \quad Z=\begin{pmatrix}
        1 & 0 & 0& \cdots  & 0 \\
        0 & \omega &0 &\cdots& 0\\
        0 & 0 & \omega^2 & \cdots & 0\\
        \vdots & \vdots & \vdots & \ddots&  \vdots\\
        0 & 0 &0&  \cdots & \omega^{N-1}
    \end{pmatrix},\label{eq:sigmatau}
\end{equation}where $\omega=e^{2\pi i/N}$. They satisfy the relations
\begin{equation}
    X^N=Z^N=1, \quad X^\dagger=X^{N-1}, \quad Z^\dagger=Z^{N-1},\quad ZX=\omega XZ.
\end{equation}The last condition is referred to as $\omega$-commutation in the literature \cite{Smith}. From them the operators $X_j= \cdots \otimes 1\otimes \underset{j \text{th}}{X} \otimes 1 \otimes \cdots$ and $Z_j=\cdots \otimes 1\otimes \underset{j \text{th}}{Z} \otimes 1 \otimes \cdots$ can be defined that act non-trivially only at the $j$th site of the chain. The shift and clock operators generate the Weyl-Heisenberg displacement operators $D_j^{(\bm{a})}=\omega^{\frac{a_1 a_2}{2}}X^{a_1}_j Z^{a_2}_j$, which form the unitary Schwinger basis \cite{SchwingerBasis} of operators in the $N$-dimensional Hilbert space. They have the properties 
\begin{equation}
    D_j^{(\bm{a})}D_j^{(\bm{b})}=\omega^{\frac{a_2b_1-a_1b_2}{2}}D^{(\bm{a}+\bm{b})}_j, \quad D_j^{(\bm{a}) \dagger} =D_j^{(-\bm{a})}.
\end{equation} The permutation operator can be expressed in term of the displacement operators as \cite{klich2024swaptransposedisplacementsstabilizer}
\begin{equation}
    P_{j,k}=\frac{1}{N}\sum_{a_1,a_2=1}^{N}D_j^{(-\bm{a})}\otimes D_k^{(\bm{a})}=\frac{1}{N}\sum_{a_1,a_2=1}^{N}\omega^{a_1a_2}X_j^{-a_1}Z_j^{-a_2}\otimes X_k^{a_1}Z_k^{a_2}.\label{eq:permutation}
\end{equation}The Schwinger basis can be transformed to and from the standard basis $(E^{(\bm{a})})_{\alpha\beta}=\delta_{a_1,\alpha}\delta_{a_2,\beta}$ by
\begin{equation}
        X^aZ^b=\sum_{n=1}^{N}\omega^{bn}E^{(n+a,n)}, \quad E^{(a,b)}= \frac{1}{N}\sum_{n=1}^{N}\omega^{-bn}X^{a-b}Z^n. \label{eq:transSchiwnger}
\end{equation}

For $N>2$, the Ising Hamiltonian \eqref{eq:Ising} generalizes to the $Z_N$-symmetric chain of von Gehlen and Rittenberg \cite{HowesKadanoffNijs,GehlenRittenberg} (in a representations with $X$ being diagonal instead of $Z$)
\begin{equation}
    H_\mathrm{vGR}= -\sum_{j}(h_{j+\frac{1}{2}}+Uh_j), \label{eq:vGRHam}
\end{equation}with
\begin{equation}
    h_{j+\frac{1}{2}}=\sum_{n=1}^{N-1}\frac{2}{1-\omega^{-n}}Z_j^{-n}Z_{j+1}^n,\quad 
    h_j=\sum_{n=1}^{N-1}\frac{2}{1-\omega^{-n}}X_j^n.
\end{equation}

Parafermions are introduced by the Fradkin-Kadanoff transformation \cite{FRADKIN19801}
\begin{equation}
    \chi_j=\left(\prod_{k=1}^{j-1}X_k\right)Z_j,\quad \psi_j=\omega^{\frac{N-1}{2}}\chi_jX_j=\omega^{\frac{N-1}{2}}\left(\prod_{k=1}^{j-1}X_k\right)Z_jX_j, \label{eq:JWpara}
\end{equation}which satisfy the relations
\begin{equation}
    \psi_j^N=\chi_j^N=1, \quad \chi_j^\dagger=\chi_j^{N-1}, \quad\psi_j^\dagger=\psi_j^{N-1}, \quad  \chi_j\psi_j=\omega\psi_j\chi_j. \label{eq:majoranaparafermion1}
\end{equation}Unlike shift and clock operators, which commute at different lattice sites, they realize the generalized Clifford algebra \cite{10.1063/1.528788, Smith}
\begin{equation}
   \chi_j\chi_k=\omega \chi_k\chi_j, \quad  \psi_j\psi_k=\omega\psi_k\psi_j,\quad \chi_j\psi_k=\omega \psi_k\chi_j,\quad \psi_j\chi_k=\omega \chi_k\psi_j,\quad \text{for } j<k, \label{eq:majoranaparafermion2}
\end{equation}due to the Klein factors in \eqref{eq:JWpara}. In terms of the parafermion operators, the Hamiltonian \eqref{eq:vGRHam} can be rewritten using $Z_j^\dagger Z_{j+1}=\omega^\frac{1-N}{2}\psi_j^\dagger \chi_{j+1}$ and $X_j=\omega^\frac{1-N}{2}\chi_j^\dagger\psi_j$ as \cite{Fendley_2012}
\begin{equation}
    H_\mathrm{vGR}=-\sum_j \sum_{n=1}^{N-1}\frac{2\omega^{\frac{n(n-N)}{2}}}{1-\omega^{-n}}\left(\psi_j^{-n}\chi_{j+1}^n+U\chi_j^{-n}\psi_j^n\right). \label{eq:clockHam}
\end{equation}

\section{The chiral Potts model}\label{sec:CPM}

The BWs of the integrable CPM was solved in the late eighties thanks to the collective wisdom and an impressive amount of hard work by some of the most insightful mathematical physicists \cite{AuMcCoyPerkTangYan,McCoyPerkTangSah,AuMcCoyPerkTang,BaxterPerkAu}. A pedagogical introduction can be found in Ref.~\cite{masterkey}. The general BWs on horizontal and vertical bonds are expressed in term of two spectral parameters $\lambda,\mu$ as \cite{AuMcCoyPerkTangYan,McCoyPerkTangSah,AuMcCoyPerkTang,BaxterPerkAu,masterkey}
\begin{equation}
     W_n(\lambda,\mu)=\prod_{j=1}^n\frac{d(\lambda)b(\mu)-a(\lambda)c(\mu)\omega^j}{b(\lambda)d(\mu)-c(\lambda)a(\mu)\omega^j},\quad 
        \overline{W}_n(\lambda,\mu)=\prod_{j=1}^n\frac{\omega a(\lambda)d(\mu)-d(\lambda)a(\mu)\omega^j}{c(\lambda)b(\mu)-b(\lambda)c(\mu)\omega^j},\label{eq:CPMBWs}
\end{equation}for adjacent $Z_N$ spins differing by $n\ne 0$ in their configurations and $W_0(\lambda,\mu)=\overline{W}_0(\lambda,\mu)=1$. The parametrization \eqref{eq:ellipara2} corresponds to $\omega=-1, n=1$ and 
\begin{equation}
    \big(a(\mu),b(\mu),c(\mu),d(\mu)\big)=\big(\Theta(iu),iH(iu),H_1(iu),\Theta_1(iu)\big).
\end{equation}

The $Z_n$ periodicity 
\begin{equation}
    W_{n+N}(\lambda,\mu)=W_n(\lambda,\mu),\quad \overline{W}_{n+N}(\lambda,\mu)=\overline{W}_n(\lambda,\mu) \label{eq:periodicity}
\end{equation}constrains the parameters along the intersection of two Fermat surfaces
\begin{equation}
    a(\mu)^N+Ub(\mu)^N=\sqrt{1-U^2}d(\mu)^N, \quad Ua(\mu)^N+b(\mu)^N=\sqrt{1-U^2}c(\mu)^N. \label{eq:Fermatcurve}
\end{equation}In addition, they satisfy the regularity
\begin{equation}
    W_n(\mu,\mu)=1, \quad  \overline{W}_n(\mu,\mu)=\frac{a(\mu)d(\mu)}{b(\mu)c(\mu)}\delta_{n,0}, \label{eq:regularity}
\end{equation}and inversion relations
\begin{equation}
    W_n(\lambda,\mu)W_n(\mu,\lambda)=1, \quad \sum_{m=0}^{N-1}\overline{W}_m(\lambda,\mu)\overline{W}_{n-m}(\mu,\lambda)=\delta_{n,0}. \label{eq:inversion}
\end{equation}The last identity follows from the convolution theorem and the discrete Fourier transform (DFT) 
\begin{equation}
    \widehat{\overline{W}}_m(\lambda,\mu)=\widehat{\overline{W}}_m(\mu,\lambda)^{-1}=\frac{1}{N}\sum_{k=1}^N\omega^{-km}\overline{W}_m(\lambda,\mu)= \prod_{j=1}^m\frac{ d(\lambda)a(\mu)-b(\lambda)c(\mu)\omega^{j-1}}{a(\lambda)d(\mu)-c(\lambda)b(\mu)\omega^{j-1}},
\end{equation}for $m\ne0$ with $\widehat{\overline{W}}_0(\lambda,\mu)=1$.

There are two interesting special cases of the CPM. The first is the superintegrable case \cite{superintegrable,Albertini,Bxtersuperintegrable,Albertinireview} that corresponds to the von Gehlen-Rittenberg (vGR) Hamiltonian. As an exception to generic CPMs, the spectrum of the vGR Hamiltonian is symmetric about zero.\footnote{Here, it should be clarified that the meaning of `chiral' in CPM is different from that of chiral symmetry in the Altland–Zirnbauer classification. Although a chiral symmetry implies spectral pairing around zero energy and zero modes, those of the vGR Hamiltonian do not result from a known chiral symmetry.} This happens if $a(0)=b(0)=1, c(0)=d(0)=\eta$, upon using the freedom of overall normalization of $a,b,c,d$ and translation of their argument, with $\eta^{2N}=(1+U)/(1-U)$. Then \eqref{eq:CPMBWs} simplifies to 
\begin{equation}
     W_n(0,\mu)=z(\mu)^n\prod_{j=1}^n\frac{1-y(\mu)\omega^j}{1-x(\mu)\omega^j}, \quad 
        \overline{W}_n(0,\mu)=z(\mu)^{-n}\prod_{j=1}^n\frac{\omega-x(\mu)\omega^j}{1-y(\mu)\omega^j},\label{eq:superintBWs}
\end{equation}and
\begin{equation}
    W_n(\mu,0)=z(\mu)^{-n}\prod_{j=1}^n\frac{1-x(\mu)\omega^j}{1-y(\mu)\omega^j}, \quad
    \overline{W}_n(\mu,0)=z(\mu)^{-n}\prod_{j=1}^n\frac{\omega x(\mu)-\omega^j}{y(\mu)-\omega^j},
\end{equation}by the substitution 
\begin{equation}
    x(\mu)=\eta\frac{a(\mu)}{d(\mu)}, \quad y(\mu)=\eta^{-1}\frac{c(\mu)}{b(\mu)}, \quad z(\mu)= \eta\frac{b(\mu)}{d(\mu)}.
\end{equation}They are subject to the constraints
\begin{equation}
    (x^N-1)(y^N-1)=U (y^N-x^N), \quad z^N(y^N-1)=x^N-1.
\end{equation}The offset and normalization $\mu$ are fixed by the initial conditions $x(0)=y(0)=z(0)=1$ and $x'(0)=y'(0)=-U z'(0)=-2U$. The latter is a consequence of differentiating \eqref{eq:Fermatcurve}
\begin{equation}
    a'(0)+Ub'(0)=(1+U)\eta^{-1}d'(0), \quad Ua'(0)+b'(0)=(1+U)\eta^{-1}c'(0).
\end{equation}The DFT
\begin{equation}
    \widehat{W}_n(0,\mu)=\prod_{j=1}^{n}\frac{\omega y(\mu)z(\mu)-x(\mu)\omega^j}{z(\mu)-\omega^j}, \quad
     \widehat{W}_n(\mu,0)=\prod_{j=1}^{n}\frac{\omega x(\mu)-y(\mu)z(\mu)\omega^j}{1-z(\mu)\omega^j},
\end{equation}and $\widehat{W}_0(0,\mu)=\widehat{W}_0(\mu,0)=1$ has the property $\partial_\mu\widehat{W}_n(0,\mu)|_{\mu=0}=-\partial_\mu\widehat{W}_n(\mu,0)|_{\mu=0}=\frac{-2}{1-\omega^{-n}}$. Together with $\partial_\mu\overline{W}'_n(0,\mu)|_{\mu=0}=-\partial_\mu\overline{W}'_n(\mu,0)|_{\mu=0}=\frac{-2U}{1-\omega^{-n}}$, the CPM BWs are related to the couplings in the vGR Hamiltonian. The subscripts in the above equations are always understood mod $N$, due to the $N$-periodicity 
\begin{equation}
    \widehat{\overline{W}}_n(\lambda,\mu)=\widehat{\overline{W}}_{n+N}(\lambda,\mu), \quad \widehat{W}_{n+N}(\lambda,\mu)=\widehat{W}_n(\lambda,\mu),
\end{equation}as well as \eqref{eq:periodicity} guaranteed by \eqref{eq:Fermatcurve}.

The other is the self-dual case corresponding to $U=1$, which reduces \eqref{eq:Fermatcurve} to 
\begin{equation}
    c(\mu)=d(\mu)=1, \quad a(\mu)=\omega^{\frac{1}{2}}b(\mu)=e^{2i\mu}
\end{equation}giving the BWs first solved by Fateev and Zamolodchikov \cite{FZ,AlertiniFZ}
\begin{equation}
     W^\mathrm{FZ}_n(\mu)=\prod_{j=1}^n\frac{\sin\left(\pi (j-\frac{1}{2})/N-\mu\right)}{\sin\left(\pi (j-\frac{1}{2})/N+\mu\right)}, \quad
        \overline{W}^\mathrm{FZ}_n(\mu)=\prod_{j=1}^n\frac{\sin\left(\pi (j-1)/N+\mu\right)}{\sin\left(\pi j/N-\mu\right)},\label{eq:FZBWs}
\end{equation}for $n\ne 0$ and $W^\mathrm{FZ}_0(\mu)=\overline{W}^\mathrm{FZ}_0(\mu)=1$. Since the duality transformation interchanges 
\begin{equation}
    W_n(\lambda,\mu)\leftrightarrow\widehat{\overline{W}}_n(\lambda,\mu), \quad \overline{W}_n(\lambda,\mu)\leftrightarrow\widehat{W}_n(\lambda,\mu), \quad U\leftrightarrow U^{-1},
\end{equation}the BWs on horizontal and vertical bonds of the FZ model are Fourier transforms of each other. Its $R$-operator and Yang-Baxter integrability has already been examined by Martins in his recent work \cite{MartinsRmat}. Except for this critical point, in general the BWs of the CPM cannot be expressed in terms of a single spectral parameter. This is due to the genus of the curve \eqref{eq:Fermatcurve} $g=N^2(N-2)+1$ being larger than one for $N>2$ \cite{Albertini}.

\section{From star-triangle relation to Yang-Baxter equation}\label{sec:STLYBE}

\begin{figure}[ht]
    \centering
    \includegraphics[width=0.25\linewidth]{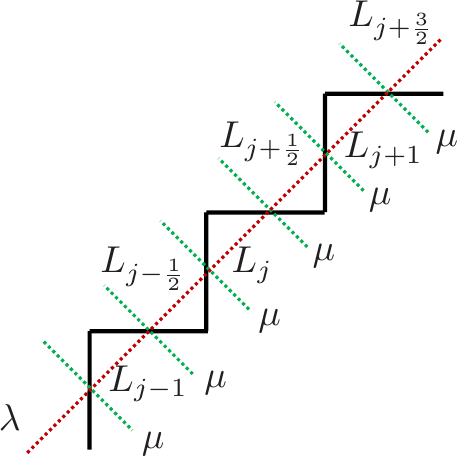}
    \caption{The diagonal-to-diagonal transfer matrix. The operator $L_j$ encodes the BW on the vertical bond in the $j$th position, and $L_{j+\frac{1}{2}}$ corresponds to the BW on horizontal bond between the $j$th and $(j+1)$ position. Red and green dotted lines denote respectively the spectral parameters $\lambda,\mu$.}
    \label{fig:transfer}
\end{figure}

Traditionally, the 1D quantum Hamiltonian and the 2D classical CPM described in the last two sections are related by the diagonal-to-diagonal transfer matrix shown in Fig.~\ref{fig:transfer}
\begin{equation}
    T(\lambda,\mu)=\cdots L_{j+1}(\lambda,\mu)L_{j+\frac{1}{2}}(\lambda,\mu)L_{j}(\lambda,\mu)L_{j-\frac{1}{2}}(\lambda,\mu)\cdots
\end{equation}
The operators $L_{j}(\mu,\nu),L_{j+\frac{1}{2}}(\mu,\nu)$ can be defined in terms of BWs $\overline{W}_n(\mu,\nu)$ and $W_n(\mu,\nu)$ on vertical and horizontal bonds respectively, which depend only on the difference $n$ of the $Z_N$ spin configurations 
\begin{equation}
\begin{split}
    L_{j}(\lambda,\mu)=&\sum_{l,m,n=1}^{N} \overline{W}_{n-m}(\lambda,\mu)E^{(l,m)}_j\otimes E_{j+1}^{(n,l)}=P_{j,j+1}\sum_{n=1}^{N}\overline{W}_n(\lambda,\mu)X_j^n,\\
    L_{j+\frac{1}{2}}(\lambda,\mu)=&\sum_{m,n=1}^{N} W_{n-m}(\lambda,\mu)E^{(m,m)}_j\otimes E^{(n,n)}_{j+1}=\sum_{n=1}^{N}\widehat{W}_n(\lambda,\mu)Z_j^{-n}\otimes Z_{j+1}^n,   
\end{split}\label{eq:L}
\end{equation}where the transformation \eqref{eq:transSchiwnger} has been used, and $\widehat{W}_n(\lambda,\mu)=\frac{1}{N}\sum_{m=1}^{N}W_m(\lambda,\mu)\omega^{-mn}$ is the DFT of $W_m(\lambda,\mu)$. 

\begin{figure}[ht]
    \centering
    \includegraphics[width=\linewidth]{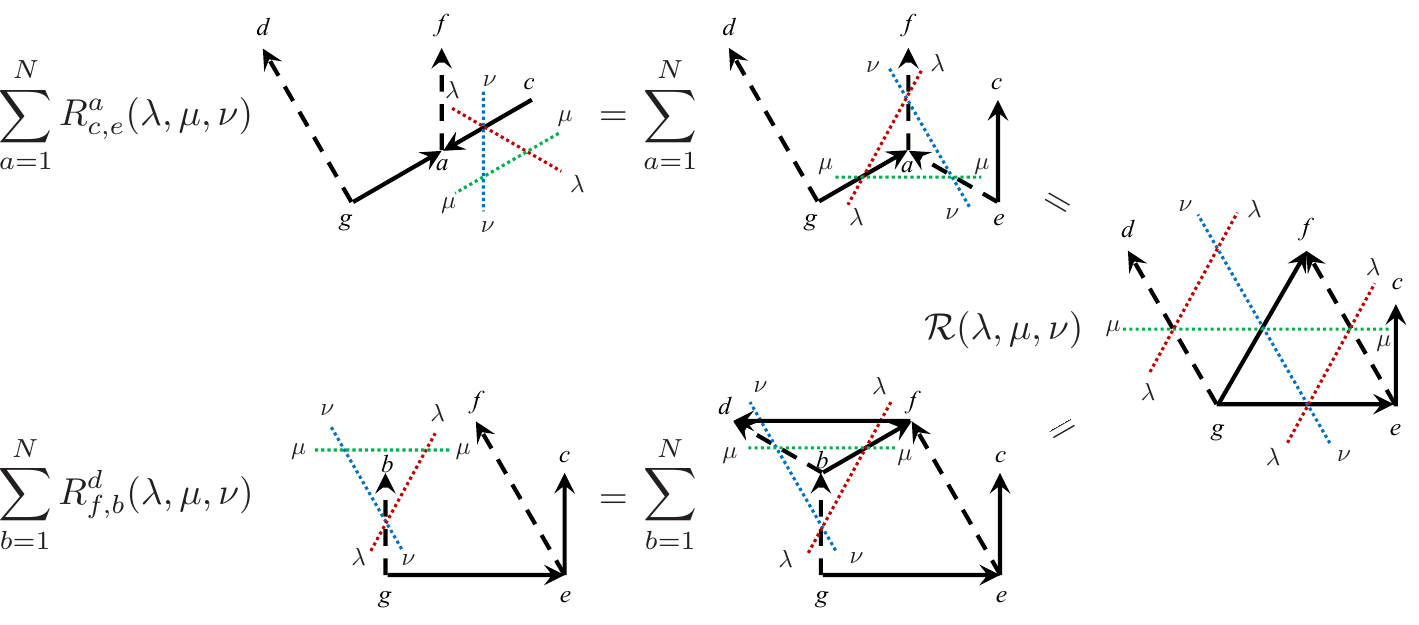}
    \caption{Graphical derivation of the YBE \eqref{eq:MartinsYBE} using the STRs \eqref{eq:STR}. Solid lines denote the BWs $W$ on horizontal bonds, dashed lines denote the BWs $\overline{W}$ on vertical bonds, and red, green, and blue dotted lines denote respectively the spectral parameters $\lambda,\mu,\nu$. The arrows on solid (resp. dashed) lines are chosen such that for $W_{a-c}(\nu,\lambda)$ (resp.~$\overline{W}_{a-c}(\nu,\lambda)$), if rotating counter-clockwise (resp. clockwise) the line between $a$ and $c$, it first aligns with the dotted line for $\nu$ before it does with the dotted line for $\lambda$, then the arrow points from $a$ towards $c$. For $W_{a-c}(\lambda,\nu)=W^{-1}_{a-c}(\nu,\lambda)$ and $\overline{W}_{a-c}(\lambda,\nu)$, the arrow points instead from $c$ to $a$.}
    \label{fig:STRYBE}
\end{figure}

Now we can construct a Lax operator
\begin{equation}
    L_{j,k}(\lambda,\mu)=\sum_{l,m,n=1}^{N}\overline{W}_{n-m}(\lambda,\mu)W_{n-l}(\lambda,\mu)E_j^{(l,m)}\otimes E_{k}^{(n,l)},
\end{equation}such that $L_{j,j+1}(\lambda,\mu)=L_{j+\frac{1}{2}}(\lambda,\mu)L_{j}(\lambda,\mu)$, to find an $R$-operator that satisfies the $RLL$ relation 
\begin{equation}
    \bm{R}_{12}(\lambda,\mu,\nu)L_{13}(\lambda,\mu)L_{23}(\lambda,\nu)=L_{23}(\lambda,\nu)L_{13}(\lambda,\mu)\bm{R}_{12}(\lambda,\mu,\nu).\label{eq:RLL}
\end{equation}In order to make the distinction between algebraic identities and operator identities more explicit, form now on bold font will be used for the $R$-operator. Plugging in the parametrization
\begin{equation}
    \bm{R}_{j,k}(\lambda,\mu,\nu) =\sum_{a,b,c,d=1}^{N}R_{a,c}^{b,d}(\lambda,\mu,\nu) E_j^{(a,b)}\otimes E_{k}^{(c,d)}, \label{eq:ansatzR}
\end{equation}and comparing the matrix elements of $E_1^{(c,d)}E_2^{(e,f)}E_3^{(g,c)}$, we have $R_{a,c}^{b,d}(\lambda,\mu,\nu)=R^{b}_{a,c}(\lambda,\mu,\nu)\delta_{a,d}$ with $R_{a,c}^b(\lambda,\mu,\nu)$ satisfying
\begin{equation}
\begin{split}
    &\sum_{a=1}^{N}R^a_{c,e}(\lambda,\mu,\nu)\overline{W}_{g-d}(\lambda,\mu)W_{g-a}(\lambda,\mu)\overline{W}_{a-f}(\lambda,\nu)W_{a-c}(\lambda,\nu)\\
    =&\sum_{b=1}^{N}R^d_{f,b}(\lambda,\mu,\nu)\overline{W}_{g-b}(\lambda,\nu)W_{g-e}(\lambda,\nu)\overline{W}_{e-f}(\lambda,\mu)W_{e-c}(\lambda,\mu)
\end{split}\label{eq:MartinsYBE}
\end{equation}for $c,d,e,f,g=1,2,\cdots, N$. In order for these equations to be guaranteed by the STRs
\begin{equation}
\begin{aligned}
    \sum_{a=1}^{N}\overline{W}_{a-f}(\lambda,\nu)W_{g-a}(\lambda,\mu)\overline{W}_{e-a}(\nu,\mu)&=\mathcal{R}(\lambda,\mu,\nu)  W_{g-f}(\nu,\mu)\overline{W}_{e-f}(\lambda,\mu) W_{g-e}(\lambda,\nu)\\
    \sum_{b=1}^{N}\overline{W}_{g-b}(\lambda,\nu)W_{b-f}(\lambda,\mu)\overline{W}_{b-d}(\nu,\mu)&=\mathcal{R}(\lambda,\mu,\nu)W_{g-f}(\nu,\mu)\overline{W}_{g-d}(\lambda,\mu)  W_{d-f}(\lambda,\nu), 
\end{aligned}\label{eq:STR}
\end{equation}we must have 
\begin{equation}
    R^b_{a,c}(\lambda,\mu,\nu) =W_{b-a}(\nu,\lambda)\overline{W}_{c-b}(\nu,\mu)W_{c-a}(\lambda,\mu). \label{eq:3pR}
\end{equation}A diagrammatic derivation of \eqref{eq:MartinsYBE} using \eqref{eq:STR} is illustrated in Fig.~\ref{fig:STRYBE}. 

\begin{figure}
    \centering
    \includegraphics[width=0.6\linewidth]{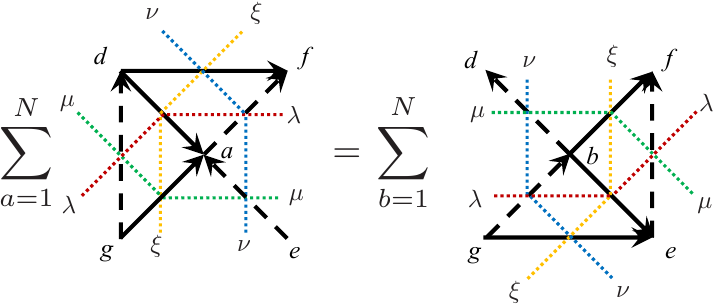}
    \caption{Diagrammatic representation of the star-star relation \eqref{eq:starstar} that underpins the YBE \eqref{eq:YBE}. Solid (resp.~dashed) lines denote the BWs $W$ on horizontal bonds (resp.~$\overline{W}$ on vertical bonds), and spectral parameters $\xi,\lambda,\mu,\nu$ are marked by yellow, red, green and blue dotted lines respectively.}
    \label{fig:decSTR}
\end{figure}

It can be easily verified that the above $R$-operator has the following properties
\begin{equation}
    \begin{split}
        \bm{R}_{j,k}(\lambda,\mu,\lambda)=&L_{j,k}(\lambda,\mu),\\
        \bm{R}_{j,k}(\lambda,\mu,\mu)=&\frac{a(\mu)d(\mu)}{b(\mu)c(\mu)}P_{j,k},\\
        \bm{R}_{k,j}(\lambda,\nu,\mu)R_{j,k}(\lambda,\mu,\nu)=&1,
    \end{split}
\end{equation}where the regularity and inversion relation results respectively from \eqref{eq:regularity} and \eqref{eq:inversion}. It only remains to check that this $R$-operator indeed satisfies the operator YBE
\begin{equation}
    \bm{R}_{12}(\kappa,\mu,\nu)\bm{R}_{13}(\xi,\mu,\lambda)\bm{R}_{23}(\kappa,\nu,\lambda)=\bm{R}_{23}(\xi,\nu,\lambda)\bm{R}_{13}(\kappa,\mu,\lambda)\bm{R}_{12}(\xi,\mu,\nu),\label{eq:YBEoperator}
\end{equation}where the rearrangement of the spectral parameters, especially the additional ones $\kappa$ and $\xi$, can be diagrammatically represented by Fig.~\ref{fig:BS} (c). Its $E_1^{(c,d)}E_2^{(e,f)}E_3^{(g,c)}$ component requires
\begin{equation}
    \sum_{a=1}^{N}R_{c,e}^a(\kappa,\mu,\nu)R_{a,g}^d(\xi,\mu,\lambda)R_{c,a}^f(\kappa,\nu,\lambda)=\sum_{b=1}^{N}R_{e,g}^b(\xi,\nu,\lambda)R_{c,e}^f(\kappa,\mu,\lambda)R_{f,b}^d(\xi,\mu,\nu). \label{eq:YBE}
\end{equation}After removing common multiplicative factors, the condition becomes
\begin{equation}
\begin{split}
    &\overline{W}_{g-d}(\lambda,\mu)W_{d-f}(\xi,\nu)\sum_{a=1}^{N}\overline{W}_{e-a}(\nu,\mu)\overline{W}_{a-f}(\lambda,\nu)W_{d-a}(\lambda,\xi)W_{g-a}(\xi,\mu)\\
    =&\overline{W}_{e-f}(\lambda,\mu)W_{g-e}(\xi,\nu)\sum_{b=1}^{N}\overline{W}_{b-d}(\nu,\mu)\overline{W}_{g-b}(\lambda,\nu)W_{b-e}(\lambda,\xi)W_{b-f}(\xi,\mu),
\end{split}\label{eq:starstar}
\end{equation}for all $d,e,f,g=1,2,\cdots, N$, as illustrated graphically in Fig.~\ref{fig:decSTR}. This is recognized as the star-star relation for the CPM. Unlike similar relations observed before for BWs that are functions of one spectral parameter and correspondingly $R$-operators of two \cite{starstar,baxter2013exactly,MartinsRmat}, here the BWs and $R$-operator depend on two and three spectral parameters respectively. The derivation of \eqref{eq:starstar} using \eqref{eq:STR} will be given in Appendix \ref{sec:STR}.

\section{Relation to the Bazhanov-Stroganov four-parameter \texorpdfstring{$R$}{R}-matrix}\label{sec:tau2}

\begin{figure}
    \centering
    \includegraphics[width=\linewidth]{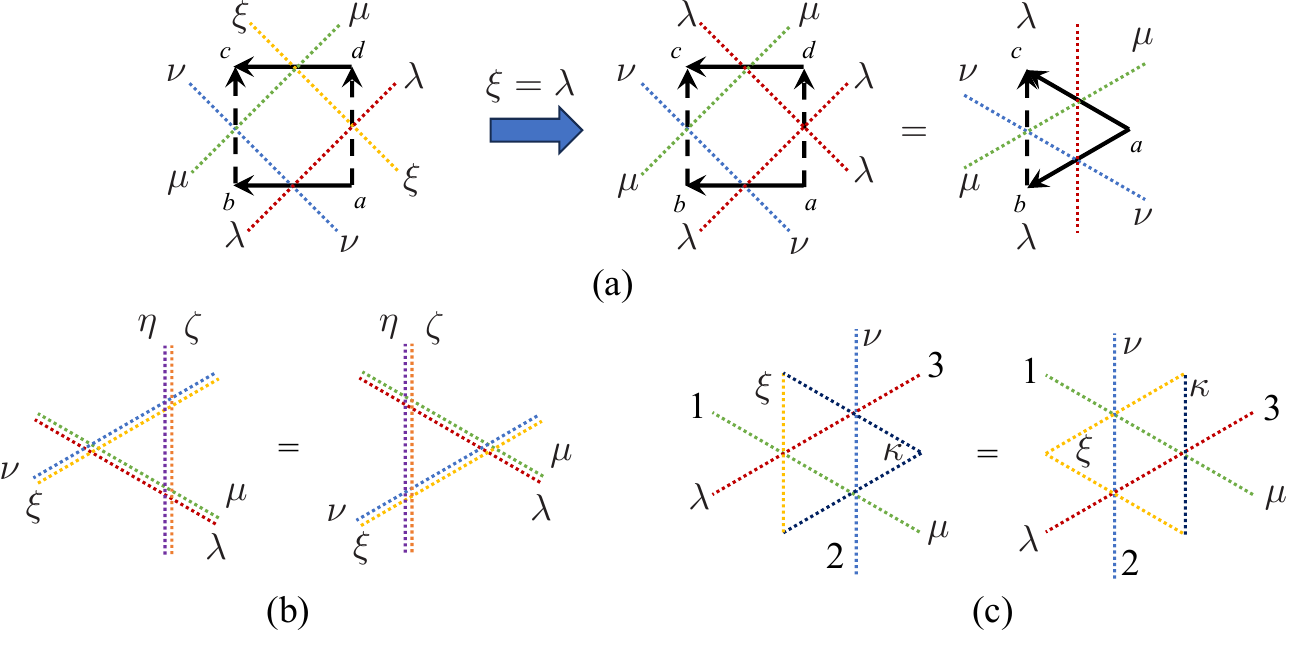}
    \caption{Diagrammatic representation of the star-star relation \eqref{eq:starstar} that underpins the YBE \eqref{eq:YBE}. Solid (resp.~dashed) lines denote the BWs $W$ on horizontal bonds (resp.~$\overline{W}$ on vertical bonds), and spectral parameters $\xi,\lambda,\mu,\nu$ are marked by yellow, red, green and blue dotted lines respectively.}
    \label{fig:BS}
\end{figure}

\begin{figure}[ht]
    \centering
    \includegraphics[width=0.8\linewidth]{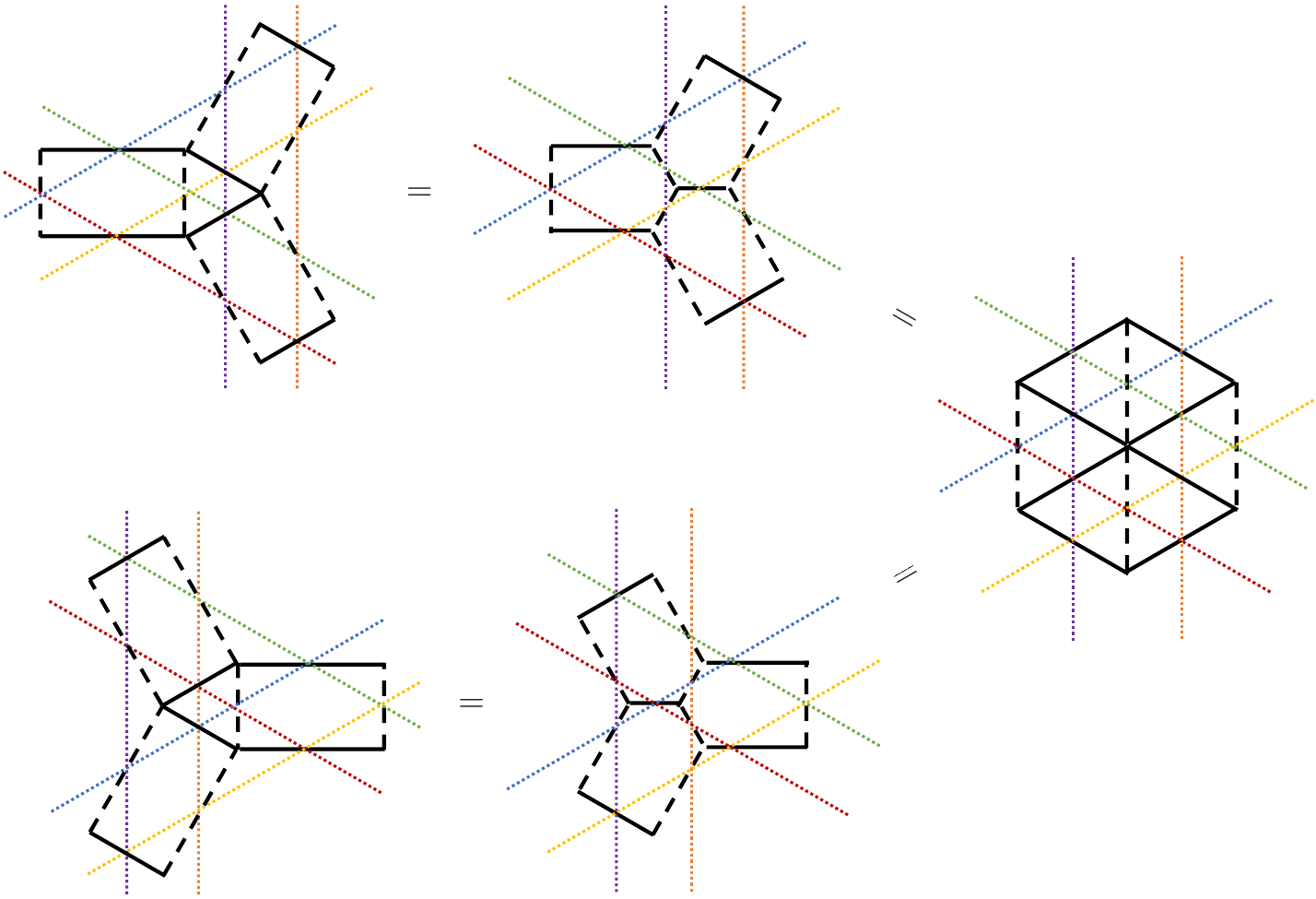}
    \caption{Diagrammatic proof of the YBE for the Bazhanov-Stroganov $R$-matrix \eqref{eq:YBE4} using the STRs \eqref{eq:STR}. The summation over the inner spins and the proportionality factors are omitted for simplicity of presentation.}
    \label{fig:YBEtau2}
\end{figure}

The components of the $R$-operator in the standard basis given by \eqref{eq:3pR} should be compared with the four-parameter $R$-matrix solved by Bazhanov and Stroganov \cite{BazhanovaStroganov,JimboMiwa}. Their $R$-matrix is solved from its intertwining property for $N$-dimensional cyclic representations of quantum group $U_q(sl_2)$ at root of unity, and is given as the product of four BWs 
\begin{equation}
    R_{b,a}^{c,d}(\lambda,\mu,\nu,\xi)=W_{b-a}(\nu,\lambda)\overline{W}_{c-b}(\nu,\mu)\overline{W}_{a-d}(\lambda,\xi)W_{c-d}(\xi,\mu),
\end{equation}as shown graphically in Fig.~\ref{fig:BS} (a). As Martins noted for the Ising case, setting $\xi=\lambda$, it simplifies to 
\begin{equation}
    R_{b,a}^{c,d}(\lambda,\mu,\nu)=\delta_{a,d}W_{b-a}(\nu,\lambda)\overline{W}_{c-b}(\nu,\mu)W_{c-a}(\lambda,\mu)=R_{a,c}^{b}(\lambda,\mu,\nu)\delta_{a,d}.
\end{equation}The four-parameter $R$-matrix satisfies the algebraic YBE 
\begin{equation}
    R_{f,g}^{b,c}(\zeta,\eta,\lambda,\mu)R_{e,b}^{d,a}(\xi,\nu,\lambda,\mu)R_{a,c}^{i,h}(\xi,\nu,\eta,\zeta)=R_{e,f}^{z,x}(\xi,\nu,\eta,\zeta)R_{x,g}^{y,h}(\xi,\nu,\lambda,\mu)R_{z,y}^{d,i}(\zeta,\eta,\lambda,\mu), \label{eq:YBE4}
\end{equation}where the repeated indices $a,b,c$ and $x,y,z$ are summed over. This algebraic YBE can be graphically represented by Fig.~\ref{fig:BS} (b), and can be derived using the STRs following the sequence of transformations shown in Fig.~\ref{fig:YBEtau2}. 

\section{Generalization to Shastry's decorated Yang-Baxter equation}\label{sec:DYBE}

Shastry's method of solving the $R$-operator for the Hubbard model, reformulated in a more general setting in Ref.~\cite{zhang2026integrablefreeinteractingfermions}, proceeds as follows. One first determines the non-interacting $R$-operator for the Hamiltonian without the onsite potential term. These non-interacting $R$-operators are then regarded as Lax operators, and auxiliary operators $X$ and $Y$ are solved from equations of the form $L_1L_2Y_{12}=X_{23}L_1L_2$. For the Hubbard and XYh models, there are two independent relations of this type, related by the multiplication with the onsite potential. One is the usual YBE for the non-interacting Hamiltonian, while the other is the so-called decorated YBE. Finally, a suitable linear combination of the $X$ (or equivalently $Y$) operators is identified as the intertwiner by imposing that it satisfies the interacting YBE for the non-relativistic $R$-operator.

To see how this strategy could be applied to the CPM, it is instructive to obtain an expression of $\bm{\check{R}}_{j,j+1}(\lambda,\mu,\nu)=P_{j,j+1}\bm{R}_{j,j+1}(\lambda,\mu,\nu)$ in the Schwinger basis
\begin{equation}
\begin{split}
     \bm{\check{R}}_{j,j+1}(\lambda,\mu,\nu)=&\sum_{a,b,c=1}^{N} R_{a,c}^b(\lambda,\mu,\nu) E_j^{(c,b)} E_{j+1}^{(a,a)}\\
     =&\frac{1}{N^2}\sum_{a,b,c,m,n=1}^{N} W_{b-a}(\nu,\lambda)\overline{W}_{c-b}(\nu,\mu)W_{c-a}(\lambda,\mu)\omega^{-bm-na}X_j^{c-b}Z_j^{m}Z_{j+1}^n\\
     =&\frac{1}{N^2}\sum_{b,k,l,m,n=1}^{N}W_{k}(\nu,\lambda)\overline{W}_{l}(\nu,\mu)W_{k+l}(\lambda,\mu)\omega^{-b(m+n)+k n}X_j^{l}Z_j^{m} Z_{j+1}^n\\
    =&\frac{1}{N}\sum_{k,l,m,n=1}^{N} \omega^{(k+l)n}W_{k}(\nu,\lambda)\overline{W}_{l}(\nu,\mu)W_{k+l}(\lambda,\mu)Z_j^{-n} Z_{j+1}^nX_j^{l}\\
    =&\sum_{n,m=1}^{N} W_{m-n}(\nu,\lambda)\overline{W}_{n}(\nu,\mu)W_{m}(\lambda,\mu)Q_{j+\frac{1}{2}}^{(m)} X_j^{n},
\end{split}\label{eq:RSchwinger}
\end{equation}where the projection operators
\begin{equation}
    Q_{j+\frac{1}{2}}^{(n)}=\frac{1}{N}\sum_{k=1}^{N}\omega^{nk}Z_j^{-k} Z_{j+1}^{k}=\sum_{m=1}^NE_j^{(m+n,m+n)}E_{j+1}^{(m,m)}, 
\end{equation}satisfy $Q_{j+\frac{1}{2}}^{(m)}Q_{j+\frac{1}{2}}^{(n)}=\delta_{m,n}Q_{j+\frac{1}{2}}^{(n)}$ and $\sum_{n=1}^{N}Q_{j+\frac{1}{2}}^{(n)}=1$. The shift operators act on them as
\begin{equation}
    X_jQ_{j+\frac{1}{2}}^{(n)}=Q_{j+\frac{1}{2}}^{(n+1)}X_j, \quad X_{j+1}Q_{j+\frac{1}{2}}^{(n)}=Q_{j+\frac{1}{2}}^{(n-1)}X_{j+1},
\end{equation}where the superscripts are understood mod $N$. One can easily check that the superintegrable Hamiltonian is reproduced by $\partial_\mu\bm{\check{R}}_{j,j+1}(0,\mu,0)|_{\mu=0}=-(h_{j+\frac{1}{2}}+Uh_j)$.

At $U=0$, the non-interacting Hamiltonian excluding the onsite potential can be written in terms of the projectors as
\begin{equation}
    h_{j+\frac{1}{2}}=\sum_{n=1}^{N-1}\sum_{k=1}^{N}\frac{2\omega^{-nk}}{1-\omega^{-n}}Q_{j+\frac{1}{2}}^{(k)}=\sum_{k=1}^{N}\left(2k-N-1\right)Q_{j+\frac{1}{2}}^{(k)},
\end{equation}where the identity $S_k=\sum_{n=1}^{N-1}\frac{2\omega^{nk}}{1-\omega^{-n}}=2k-N-1$ has been used.\footnote{It can be easily seen by noticing $S_N=\sum_{n=1}^{N-1}(\frac{1}{1-\omega^{-n}}+\frac{1}{1-\omega^{n}})=N-1$ and $S_{k}-S_{k+1}=2\sum_{n=1}^{N-1}\omega^{-nk}=2N\delta_{k,0}-2$.} They mutually commute on different lattice sites, and $L_{j+\frac{1}{2}}$ in \eqref{eq:L} can be identified with the non-interacting $R$-operator for $h_{j+\frac{1}{2}}$ 
\begin{equation}
\begin{split}
    L_{j+\frac{1}{2}}(\lambda,\mu)=\sum_{n=1}^{N}\widehat{W}_{n}(\lambda,\mu)Z_j^{-n}\otimes Z_{j+1}^{n}=\sum_{k=1}^NW_{-k}(\lambda,\mu)Q_{j+\frac{1}{2}}^{(k)}.
    \end{split}   \label{eq:noninteractingR}
\end{equation}Due to the diagonal nature of these matrices, equations analogous to Shastry's DYBE 
\begin{equation}
    L_{j-\frac{1}{2}}(\lambda,\nu)L_{j+\frac{1}{2}}(\lambda,\mu)\bm{Y}_{j-1,j}(\lambda,\mu,\nu)=\bm{X}_{j,j+1}(\lambda,\mu,\nu)L_{j-\frac{1}{2}}(\lambda,\mu)L_{j+\frac{1}{2}}(\lambda,\nu) \label{eq:RLLdecorated}
\end{equation}can only be satisfied if the unknown operators are of the form
\begin{equation*}
\begin{split}
    \bm{X}_{j,j+1}(\lambda,\mu,\nu)=&\sum_{a,b,c=1}^{N}  X_{a,c}^{b}(\lambda,\mu,\nu)E_j^{(a,b)} E_{j+1}^{(c,c)},\\
    \bm{Y}_{j,j+1}(\lambda,\mu,\nu)=&\sum_{a,b,d=1}^{N}Y_{a,d}^{b}(\lambda,\mu,\nu)E_{j}^{(d,d)} E_{j+1}^{(a,b)}.
\end{split}
\end{equation*}In this parameterization, the coefficients of $E_{j-1}^{(c,c)}E_j^{(a,b)}E_{j+1}^{(d,d)}$ in the equation becomes
\begin{equation}
    W_{b-c}(\nu,\lambda)W_{a-c}(\lambda,\mu)Y_{a,d}^b(\lambda,\mu,\nu)=X_{a,c}^b(\lambda,\mu,\nu)W_{d-b}(\lambda,\mu)W_{d-a}(\nu,\lambda).
\end{equation}
Generic solutions are given by
\begin{equation}
\begin{split}
    X_{a,c}^{b}(\lambda,\mu,\nu)=&W_{b-c}(\nu,\lambda)W_{a-c}(\lambda,\mu)f_{a,b}(\lambda,\mu,\nu), \\
    Y_{a,d}^{b}(\lambda,\mu,\nu)=&W_{d-a}(\nu,\lambda)W_{d-b}(\lambda,\mu)f_{a,b}(\lambda,\mu,\nu),
\end{split}\label{eq:XVcoeff}
\end{equation}for some arbitrary function $f_{a,b}(\lambda,\mu,\nu)$. According to \eqref{eq:RSchwinger}, choosing $f_{a,b}(\lambda,\mu,\nu)=\overline{W}_{a-b}(\nu,\mu)$ gives the interacting $R$-operator that satisfies the YBE \eqref{eq:YBEoperator}.

Therefore, \eqref{eq:RLLdecorated} with 
\begin{equation}
    \begin{split}
        \bm{X}_{j,j+1}(\lambda,\mu,\nu)=&\sum_{a,b,c=1}^{N}W_{b-c}(\nu,\lambda)W_{a-c}(\lambda,\mu)f_{a-b}(\nu,\mu)E_j^{(a,b)} E_{j+1}^{(c,c)}\\
        =&\sum_{m,n=1}^{N} W_{m-n}(\nu,\lambda)W_{m}(\lambda,\mu)f_{n}(\nu,\mu)Q_{j+\frac{1}{2}}^{(m)}X_j^{n}, \\
        \bm{Y}_{j,j+1}(\lambda,\mu,\nu)=&\sum_{a,b,c=1}^{N}W_{d-a}(\nu,\lambda)W_{d-b}(\lambda,\mu)f_{a-b}(\nu,\mu)E_j^{(d,d)} E_{j+1}^{(a,b)}\\
        =&\sum_{m,n=1}^{N} W_{m-n}(\nu,\lambda)W_{m}(\lambda,\mu)f_{n}(\nu,\mu)X_j^{n}Q_{j+\frac{1}{2}}^{(m)}
    \end{split}
\end{equation}can be considered as generalizations to Shastry's decorated YBE for the CPM and integrable parafermions. In contrast to the fermionic case \cite{FermiR}, where the $N=2$ independent equations are just the YBE for non-interacting $R$-operator and its `time-reversal' partner generated by the onsite potential in the local Hamiltonian, in the parafermionic case the $X$ and $Y$ operators appear as linear combinations of different powers of the onsite potential, with coefficients encoded in an arbitrary function $f_n(\nu,\mu)$.

\section{Conclusion}\label{sec:conclusion}

The main result of this paper is an $R$-operator, i.e. an integral form for the local Hamiltonian of the vGR parafermion chain. This $R$-operator provides the unitary quantum gates in an integrable quantum circuit realization of the model, and satisfies a new type of YBE implied by the STR of its classical dual---the CPM. The construction simultaneously generalizes Martins' embedding of edge models with single-parameter BWs into vertex models, and unifies the Bazhanov-Stroganov $R$-matrix construction with Shastry's decorated YBE. Taken together, these results offer substantive evidence toward resolving the long-standing question of whether the YBE underlies all integrable two-dimensional statistical mechanical models and their dual one-dimensional quantum Hamiltonians, within a framework that relates Onsager’s STR to the YBE. In particular, the analysis clarifies the origin of non-relativistic $R$-operators: They may arise either from the presence of onsite potentials in the Hamiltonian in addition to nearest-neighbor interactions, and/or from the higher-genus rapidity curve parameterized by the spectral parameters. For the CPM, both mechanisms are operative, leading to an $R$-operator that depends on three spectral parameters. 

Previously, the quantum–classical correspondence between exactly solvable 2D statistical models and 1D spin chains was understood primarily in the strongly anisotropic limit of the transfer matrix. The discrete Fourier transform between the Schwinger basis built from clock and shift operators and the standard basis employed here yields a sharper and more general relation between the BWs of horizontal and vertical bonds and local Hamiltonian operators, valid at arbitrary interaction strength. This viewpoint also helps explain why models admitting a duality transformation between order and disorder operators, such as the Ising model and the CPM, occupy a particularly distinguished place among integrable systems. From this perspective, it is natural to regard the YBE as the more fundamental integrability condition, with the STR emerging as a more restrictive condition realized only in special cases like the Ising and CPM. Nevertheless, it is appealing to search for generalized STRs in models already known to satisfy the YBE. In this respect, the edge-vertex correspondence constructed by Bazhanov and Sergeev \cite{BAZHANOV2023116055} offers a concrete and promising direction.

A key advantage of working with the $R$-operator rather than $R$-matrices is that it shifts the focus integrability from the transfer matrix of a classical statistical model to the unitary evolution operator in a quantum circuit. This opens up new research avenues, including non-Hermitian Hamiltonians generated from the same $R$-operator of an integrable Hermitian one. It will be worthwhile to investigate how these Hamiltonians are related to the higher Hamiltonians of relativistic $R$-operators, and whether derivatives of three-parameter $R$-operators generate a broader class of integrable Hamiltonians than other non-relativistic constructions.

Although the present constructions of $R$-operator for the vGR chain extends a special case of Shastry's approach, namely the Ising/Majorana chain, the extension to parafermionic analogues of the more general XYh and Hubbard models remains an open problem. Such extensions would naturally involve the Fock parafermions introduced in Ref.~\cite{PhysRevA.89.012328} and reviewed in Appendix \ref{sec:Dirac}. One possible direction is to exploit the properties of generalized hyperbolic functions \cite{GHF,Truong_1990} to construct an $R$-operator for the `non-interacting' Fock parafermions, using projection operators analogous to those defined in Sec.~\ref{sec:DYBE}. It is conceivable, however, that there are fundamental obstructions to this program: Parafermions may be intrinsically interacting objects \cite{Fendley_2014}, in which case a direct non-interacting limit in the spirit of Shastry’s construction might simply not exist. If so, a more realistic generalization of the decorated YBE approach might be found instead in non-Hermitian settings, such as the $Z_N$ clock model \cite{Asimple,Fendley_2014}. Exploring these directions appears to be a natural next step in developing and classifying variants of YBEs in parafermionic and non-Hermitian quantum systems.

\section*{Acknowledgements}
I thank Christopher Ekman, Paul Fendley, Israel Klich and Viktor Svensson for valuable discussions.

\begin{appendix}
\numberwithin{equation}{section}

\section{Elliptic functions} \label{sec:elliptic}

Jacobi's theta functions used in Ref.~\cite{PhysRev.65.117} (and in their modern notations) are defined by 
\begin{equation}
    \begin{split}
        H(u)&=\theta_1\left(\frac{u}{2I}\right)=\vartheta_1\left(\frac{\pi u}{2I}\right)=2q^\frac{1}{4}\sin\frac{\pi u}{2I}\prod_{n=1}^\infty\left(1-2q^{2n}\cos\frac{\pi u}{I}+q^{4n}\right)(1-q^{2n}),\\ 
        H_1(u)&=\theta_2\left(\frac{u}{2I}\right)=\vartheta_2\left(\frac{\pi u}{2I}\right)=2q^\frac{1}{4}\cos\frac{\pi u}{2I}\prod_{n=1}^\infty\left(1+2q^{2n}\cos\frac{\pi u}{I}+q^{4n}\right)(1-q^{2n}),\\ 
        \Theta_1(u)&=\theta_3\left(\frac{u}{2I}\right)=\vartheta_3\left(\frac{\pi u}{2I}\right)=\prod_{n=1}^\infty\left(1-2q^{2n-1}\cos\frac{\pi u}{I}+q^{4n-2}\right)(1-q^{2n}),\\
        \Theta(u)&=\theta_4\left(\frac{u}{2I}\right)=\vartheta_4\left(\frac{\pi u}{2I}\right)=\prod_{n=1}^\infty\left(1+2q^{2n-1}\cos\frac{\pi u}{I}+q^{4n-2}\right)(1-q^{2n}),
    \end{split}
\end{equation}where 
\begin{equation}
    I=I(\kappa)=\int_0^1\frac{dt}{\sqrt{(1-t^2)(1-\kappa^2t^2)}}=\int_0^\frac{\pi}{2}\frac{d\theta}{\sqrt{1-\kappa^2\sin^2\theta}}
\end{equation}is the complete elliptic integral of the first kind of modulus $\kappa=(H_1(0)/\Theta_1(0))^2$, and the nome $q=e^{-\pi I'/I}$, with $I'=I(\kappa')$. 
The quasi-periodicity of the theta functions implies that the Jacobian elliptic functions
\begin{equation}
    \sn u=\frac{H(u)}{\sqrt{\kappa}\Theta(u)}, \quad \cn u=\frac{\sqrt{\kappa'}H_1(u)}{\sqrt{\kappa}\Theta(u)}, \quad \dn u =\frac{\sqrt{\kappa'}\Theta_1(u)}{\Theta(u)},
\end{equation}with the conjugate modulus $\kappa'=\sqrt{1-\kappa^2}=(\Theta(0)/\Theta_1(0))^2$ are doubly periodic
\begin{equation}
    \begin{split}
        \sn(u+2I)=-\sn u, \quad \cn(u+2I)=-\cn u, \quad \dn(u+2I)=\dn u, \\
        \sn(u+2iI')=\sn u, \quad \cn(u+2iI')=-\cn u, \quad \dn(u+2iI')=-\dn u, 
    \end{split}
\end{equation}and satisfy the relations
\begin{equation}
    \sn^2u+\cn^2 u=1, \quad \kappa^2\sn^2u+\dn^2 u=1.
\end{equation}The reciprocals of these functions are named by reversing the order of the two letters in their names
\begin{equation}
    \ns(u)=\frac{1}{sn(u)}, \quad \nc(u)= \frac{1}{\cn(u)},\quad \nd(u)=\frac{1}{\dn(u)}.
\end{equation}Similarly, the ratios between the three primary functions are labeled by the first letter of the numerator followed by the first letter of the denominator
\begin{equation}
    \begin{split}
        \mathrm{sc}(u)=\frac{\sn(u)}{\cn(u)}, \quad \sd(u)=\frac{\sn(u)}{\dn(u)}, \quad \dc(u)=\frac{\dn(u)}{\cn(u)}, \\
        \cs(u)=\frac{\cn(u)}{\sn(u)}, \quad \ds(u)=\frac{\dn(u)}{\sn(u)}, \quad \cd(u)=\frac{\cn(u)}{\dn(u)}. 
    \end{split}
\end{equation}The Jacobi amplitude $\am(u)$ is defined by $\sn u=\sin \am(u)$, such that $u$ can be expressed in terms of the incomplete elliptic integral of the first kind as
\begin{equation}
    u=\int_0^{\sn u}\frac{dt}{\sqrt{(1-t^2)(1-\kappa^2t^2)}}=\int_0^{\am(u)}\frac{d\theta}{\sqrt{1-\kappa^2\sin^2\theta}}. \label{eq:firstkind}
\end{equation}We will also need the derivatives of the elliptic functions
\begin{equation}
    \sn'(u)=\cn(u)\dn(u), \quad \cn'(u)=-\sn(u)\dn(u), \quad \dn'(u)=-\kappa\sn(u)\cn(u).
\end{equation}The addition formulas
\begin{equation}
    \begin{split}
        \sn(u\pm v)&=\frac{\sn(u)\cn(v)\dn(v)\pm\sn(v)\cn(u)\dn(u)}{1-\kappa^2\sn^2(u)\sn^2(v)},\\
        \cn(u\pm v)&=\frac{\cn(u)\cn(v)\mp\sn(u)\sn(v)\dn(u)\dn(v)}{1-\kappa^2\sn^2(u)\sn^2(v)},\\
        \dn(u\pm v)&=\frac{\dn(u)\dn(v)\mp\kappa^2\sn(u)\cn(u)\sn(v)\cn(v)}{1-\kappa^2\sn^2(u)\sn^2(v)}\\ 
    \end{split}
\end{equation}are used in Sec.~\ref{sec:IM} to show that the BWs of the Ising model indeed depend only on the difference of the two spectral parameters, as an exception to the CPM with general $N$.

So far we have been omitting the modulus $\kappa$, but the the value of elliptic functions with a purely imaginary argument is related to those of conjugate modulus $\kappa'$ by
\begin{equation}
\begin{split}
    \am(iu,\kappa)=i\am(u,\kappa'), \quad \sn(iu,\kappa)=i\mathrm{sc}(u,\kappa'), \\
    \cn(iu,\kappa)=\nc(u,\kappa'), \quad \dn(iu,\kappa)=\dc(u,\kappa').
\end{split}
\end{equation}More detailed accounts and proofs of these properties can be found in Chapter 15 of Ref.~\cite{baxter2013exactly}.

\section{Star-triangle and star-star relations for the chiral Potts model}
\label{sec:STR}

For completeness, I first sketch an elementary way to see that \eqref{eq:CPMBWs} satisfies \eqref{eq:STR} as detailed in Ref.~\cite{BaxterPerkAu,masterkey,10.1093/acprof:oso/9780199556632.001.0001}. A modern proof using hypergeometric series was later given in \cite{doi:10.1142/9789812776358_0001}. Define $V_{m,k}(\nu,\mu,\lambda)=\overline{W}_k(\nu,\mu)W_{k+m}(\lambda,\mu)$ and its DFT
\begin{equation}
    \widehat{V}_{m,n}(\nu,\mu,\lambda)=\frac{1}{N}\sum_{k=1}^N \omega^{-nk}V_{m,k}(\nu,\mu,\lambda)=\frac{1}{N}\sum_{k=1}^N \omega^{-nk}\overline{W}_k(\nu,\mu)W_{k+m}(\lambda,\mu), 
\end{equation}
such that the DFT of both equations in \eqref{eq:STR} are of the form
\begin{equation}
    \widehat{V}_{m,n}(\nu,\mu,\lambda)\widehat{\overline{W}}_{n}(\lambda,\nu)= \mathcal{R}(\lambda,\mu,\nu)W_{m}(\lambda,\nu)\widehat{V}_{m,n}(\lambda,\mu,\nu).\label{eq:STRDFT}
\end{equation}In order to verify this identity, we first obtain two recurrence relations of $V_{m,n}(\mu,\nu)$ using \eqref{eq:CPMBWs}
\begin{equation}
\begin{split}    
    \frac{V_{m-1,k}(\nu,\mu,\lambda)}{V_{m,k-1}(\nu,\mu,\lambda)}=&\frac{\overline{W}_k(\nu,\mu)}{\overline{W}_{k-1}(\nu,\mu)}=\frac{\omega a(\nu)d(\mu)-d(\nu)a(\mu)\omega^k}{c(\nu)b(\mu)-b(\nu)c(\mu)\omega^k},\\
    \frac{V_{m,k}(\nu,\mu,\lambda)}{V_{m-1,k}(\nu,\mu,\lambda)}=&\frac{W_{k+m}(\lambda,\mu)}{W_{k+m-1}(\lambda,\mu)}=\frac{d(\lambda)b(\mu)-a(\lambda)c(\mu)\omega^{k+m}}{b(\lambda)d(\mu)-c(\lambda)a(\mu)\omega^{k+m}},
\end{split}
\end{equation}Omitting the common spectral parameters of $V(\nu,\mu,\lambda)$, their DFTs are given by
\begin{equation}
    \begin{split}
     a(\nu)d(\mu)\widehat{V}_{m,n}-\omega^{n-1}c(\nu)b(\mu)\widehat{V}_{m-1,n}&=d(\nu)a(\mu)\widehat{V}_{m,n-1}-\omega^{n-1}b(\nu)c(\mu)\widehat{V}_{m-1,n-1},\\
        b(\lambda)d(\mu)\widehat{V}_{m,n}-d(\lambda)b(\mu)\widehat{V}_{m-1,n}&=\omega^{m}c(\lambda)a(\mu)\widehat{V}_{m,n-1}-\omega^{m}a(\lambda)c(\mu)\widehat{V}_{m-1,n-1}.
    \end{split} \label{eq:Vrecurr}
\end{equation}
Eliminating $\widehat{V}_{m-1,n}$ and $\widehat{V}_{m,n-1}$ from the above, we get 
\begin{equation}
\begin{split}
    a(\mu)\bigl[d(\lambda)d(\nu)-\omega^{m+n-1}c(\nu)c(\lambda)\bigr]\widehat{V}_{m,n-1}=&d(\mu)\bigl[d(\lambda)a(\nu)-\omega^{n-1}c(\nu)b(\lambda)\bigr]\widehat{V}_{m,n}\\
    &+\omega^{n-1}c(\mu)\bigl[d(\lambda)b(\nu)-\omega^{m}c(\nu)a(\lambda)\bigr]\widehat{V}_{m-1,n-1},\\
    b(\mu)\bigl[d(\nu)d(\lambda)-\omega^{m+n-1}c(\lambda)c(\nu)\bigr]\widehat{V}_{m-1,n}=&d(\mu)\bigl[d(\nu)b(\lambda)-\omega^{m}c(\lambda)a(\nu)\bigr]\widehat{V}_{m,n}\\
    &+\omega^{m}c(\mu)\bigl[d(\nu)a(\lambda)-\omega^{n-1}c(\lambda)b(\nu)\bigr]\widehat{V}_{m-1,n-1}.
\end{split}\label{eq:recurrV}
\end{equation}Now we can rename the LHS of \eqref{eq:STRDFT} by $\overline{Y}_{m,n}= \widehat{V}_{m,n}(\nu,\mu,\lambda)\widehat{\overline{W}}_{n}(\lambda,\nu)$, such that \eqref{eq:recurrV} become equivalent to
\begin{equation}
    \begin{split}
        a(\mu)\bigl[d(\lambda)d(\nu)-\omega^{m+n-1}c(\nu)c(\lambda)\bigr]\overline{Y}_{m,n-1}=&d(\mu)\bigl[a(\lambda)d(\nu)-\omega^{n-1}b(\nu)c(\lambda)\bigr]\overline{Y}_{m,n}\\
    &+\omega^{n-1}c(\mu)\bigl[d(\lambda)b(\nu)-\omega^{m}c(\nu)a(\lambda)\bigr]\overline{Y}_{m-1,n-1},\\
    b(\mu)\bigl[d(\nu)d(\lambda)-\omega^{m+n-1}c(\lambda)c(\nu)\bigr]\overline{Y}_{m-1,n}=&d(\mu)\bigl[d(\nu)b(\lambda)-\omega^{m}c(\lambda)a(\nu)\bigr]\overline{Y}_{m,n}\\
    &+\omega^{m}c(\mu)\bigl[a(\nu)d(\lambda)-\omega^{n-1}b(\lambda)c(\nu)\bigr]\overline{Y}_{m-1,n-1}.
    \end{split}
\end{equation}Likewise the recurrence relations for $V(\lambda,\mu,\nu)$ with $\lambda,\nu$ interchanged in \eqref{eq:Vrecurr} result in identical recurrence relations for $Y_{m,n}=W_{m}(\lambda,\nu)\widehat{V}_{m,n}(\lambda,\mu,\nu)$ which appears on the RHS of \eqref{eq:STRDFT}. 

The proportionality factor $\mathcal{R}(\lambda,\mu,\nu)$ can be determined by the matrix form of the STR \cite{MatveevSmirnov}. Define the diagonal matrices $\bm{W}(\mu,\nu)$ and cyclic matrix $\overline{\bm{W}}(\mu,\nu)$ with their $(a,b)$ entries being
\begin{equation}
    \bm{W}_{a,b}(\mu,\nu)=W_a(\mu,\nu)\delta_{a,b}, \quad \overline{\bm{W}}_{a,b}(\mu,\nu)=\overline{W}_{a-b}(\mu,\nu),
\end{equation}such that the STR can be expressed as the matrix identity
\begin{equation}
    \overline{\bm{W}}(\lambda,\nu)\bm{W}(\lambda,\mu)\overline{\bm{W}}(\nu,\mu)=\mathcal{R}(\lambda,\mu,\nu) \bm{W}(\nu,\mu)  \overline{\bm{W}}(\lambda,\mu)\bm{W}(\lambda,\nu).
\end{equation}The determinant of the matrix identity gives 
\begin{equation}
    \mathcal{R}(\lambda,\mu,\nu)=\frac{f(\lambda,\nu)f(\nu,\mu)}{f(\lambda,\mu)}, 
\end{equation}where 
\begin{equation}
    f(\lambda,\mu)=\left(\frac{\det\overline{\bm{W}}(\lambda,\mu)}{\det\bm{W}(\lambda,\mu)}\right)^\frac{1}{N}=N\left(\prod_{n=1}^N\frac{\widehat{\overline{W}}_n(\lambda,\mu)}{W_n(\lambda,\mu)}\right)^\frac{1}{N}.
\end{equation}

\begin{figure}[ht]
    \centering
    \includegraphics[width=\linewidth]{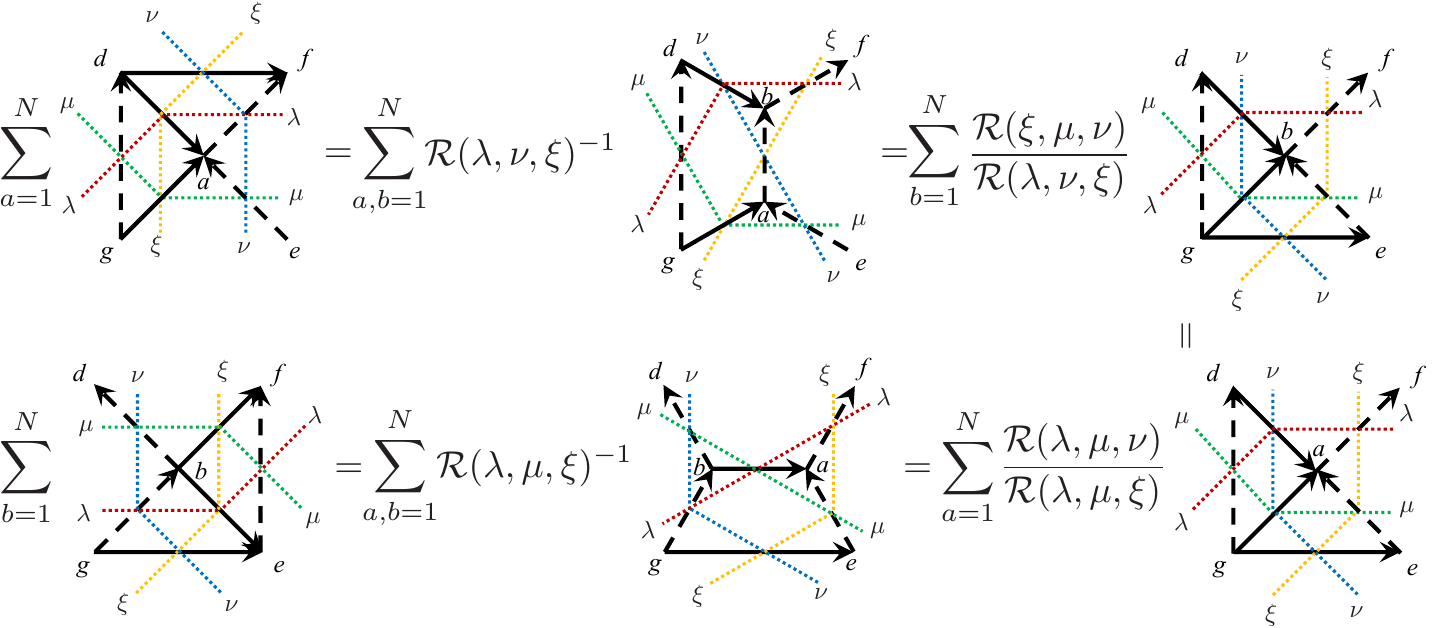}
    \caption{Diagrammatic proof of the star-star relation \eqref{eq:starstar} using the STRs \eqref{eq:STR}. }
    \label{fig:STRSRR}
\end{figure}
The star-star relation now follows readily from the sequences of star-triangle transformations shown in Fig.~\ref{fig:STRSRR} and the identity
\begin{equation}
    \frac{\mathcal{R}(\xi,\mu,\nu)}{\mathcal{R}(\lambda,\nu,\xi)}=\frac{\mathcal{R}(\lambda,\mu,\nu)}{\mathcal{R}(\lambda,\mu,\xi)}=\frac{f(\lambda,\nu)f(\nu,\mu)}{f(\lambda,\xi)f(\xi,\mu)}.
\end{equation}

\section{Dirac parafermions}\label{sec:Fock}

The parafermions introduced in Sec.~\ref{sec:ZNparafermion} with $\chi_j^N=\psi_j^N=1$ generalize the Majorana fermions with $a_j^2=b_j^2=1$ of Sec.~\ref{sec:IM}. They were named in Ref.~\cite{PhysRevA.89.012328} `Weyl parafermions', after the Weyl algebra that they realize. Personally, I find the term confusing, not the least because of the existing notion of Weyl fermions, which is a different decomposition of Dirac fermions from Majoranas. Here, I suggest calling them `Majorana parafermions' instead, in contrast to the `Dirac parafermions' that we will look for in this section.\footnote{Alternatively they could be called respectively real and complex parafermions, but since Majorana parafermion operators are already non-Hermitian and represented by complex matrices, this seems a less preferable choice.} The desired property of Dirac parafermions should instead generalize the Pauli exclusion $c_j^2=c_j^{\dagger 2}=0$ of Dirac fermions constructed from Majoranas as
\begin{equation}
    c_j=\frac{a_j+ib_j}{2}, \quad c_j^\dagger=\frac{a_j-ib_j}{2}.
\end{equation}

\subsection{From \texorpdfstring{$N$-Clifford to $N$-Grassmann}{N-Clifford to N-Grassmann} algebra}

A natural parafermionic generalization to Dirac fermions is specified by the $N$-Grassmann algebra obtained from the $N$-Clifford algebra \cite{10.1063/1.526853}, represented by
\begin{equation}
    c_j^{(\alpha)}=\frac{\chi_j+\omega^\alpha\psi_j}{2}, \quad  \alpha=\frac{1}{2},\frac{3}{2},\cdots,N-\frac{1}{2}.
\end{equation}Using $\chi^N_j=\psi^N_j=1$, $\omega^\frac{N}{2}=-1$, and the $\omega$-commutation relation, we have
\begin{equation}
\begin{split}
    \left(c_j^{(\alpha)}\right)^N= &\sum_{k=1}^{N-1}\omega^{k\alpha}\!\!\!\!\!\!\!\!\sum_{\substack{m_1,\cdots,m_{k}\ge 0\\ \sum_{l=1}^k m_l\le N-k}}\!\!\!\!\chi_j^{m_k}\psi_j\chi_j^{m_{k-1}}\psi_j\cdots\chi_j^{m_{1}}\psi_j\chi_j^{N-k-\sum_{l=1}^k m_l}\\
    =&\sum_{k=1}^{N-1}\omega^{k\alpha}\!\!\!\!\!\!\!\!\sum_{\substack{m_1,\cdots,m_{k}\ge 0\\ \sum_{l=1}^k m_l\le N-k}}\!\!\!\!\omega^{\sum_{l=1}^k lm_l}\psi_j^k\chi_j^{N-k}\\
    =&\sum_{k=1}^{N-1}\omega^{k\alpha}\binom{N}{k}_\omega\psi_j^k\chi_j^{N-k},
\end{split}
\end{equation}with the Gaussian binomial coefficient
\begin{equation}
    \binom{N}{k}_\omega=\frac{[N]_\omega!}{[k]_\omega![N-k]_\omega!}=\frac{\prod_{l=0}^{k-1}[N-l]_\omega}{\prod_{l=1}^{k}[l]_\omega}=\frac{\prod_{l=0}^{k-1}(1-\omega^{N-l})}{\prod_{l=1}^k(1-\omega^l)}=0,
\end{equation}where $[k]_q=\sum_{l=0}^{k-1}q^l$ is the $q$-number.

They carry the $\omega$-commutation relation of Majorana parafermions $c_j^{(\alpha)}c_k^{(\beta)}=\omega c_k^{(\beta)}c_j^{(\alpha)}$ for $j<k$. But for operators acting on the same lattice site, they obey $c_j^{(\alpha)}c_j^{(\beta)}= c_j^{(\beta+1)}c_j^{(\alpha-1)}$. Hermitian conjugation is given by $c_j^{(\alpha) \dagger}=c_j^{(-\alpha)}$. However, they do not satisfy the parafermionic generalization to the anticommutation relation to be introduced in the next subsection.

\subsection{Fock Parafermions} \label{sec:Dirac}

Parafermionic operators that obey the generalized canonical anticommutation relation (gCAR)
\begin{equation}
    C_j^{\dagger n} C_j^n + C_j^{N-n} C_j^{\dagger N-n}=1, \quad \text{for }\ n=1,2,.\cdots, N-1 \label{eq:anticomm}
\end{equation}were discovered in Ref.~\cite{PhysRevA.89.012328}, where they were called Fock parafermions. This relation is crucial for defining the Fock space of parafermions and normal ordering of operators in this space. In the case of $N=2$, this becomes $n_j+(1-n_j)=1$, where $n_j^2=n_j$ and $(1-n_j)^2=1-n_j$ are two projection operators onto the subspace of the $j$th fermion being occupied and unoccupied. An intuitive interpretation of \eqref{eq:anticomm} is thus that the $j$th parafermion state must be occupied by either at least $n$ parafermions, or at most $n-1$ parafermions. Therefore, the parafermion number operator that counts the number of parafermions occupying the $j$th state is given by
\begin{equation}
    N_j=\sum_{n=1}^{N-1}C_j^{\dagger n}C_j^n
    =\begin{pmatrix}
        0 & 0 & 0& \cdots  & 0 \\
        0 & 1 &0 &\cdots& 0\\
        0 & 0 & 2 & \cdots & 0\\
        \vdots & \vdots & \vdots & \ddots&  \vdots\\
        0 & 0 &0&  \cdots & N-1
    \end{pmatrix}, \label{eq:numdef}
\end{equation}since only the first $n$ terms in the sum do not annihilate a state occupied by $n$ parafermions. The anticommutativity \eqref{eq:anticomm} and telescopic cancellations ensure that
\begin{equation}
    [N_j,C_j^\dagger]=C_j^\dagger, \quad [N_j,C_j]=-C_j. 
    \label{eq:numcomm}
\end{equation}

In order to relate the Fock parafermion operators $C_j, C_j^\dagger$ to their Majorana counterparts $\chi, \psi$, the hard-core boson operators 
\begin{equation}
    B=\begin{pmatrix}
        0 & 1 & 0& \cdots  & 0 \\
        0 & 0 & 1 &\cdots& 0\\   
        \vdots & \vdots & \vdots & \ddots&  \vdots\\
         0 & 0 & 0 & \cdots & 1\\
        0 & 0 &0&  \cdots & 0
    \end{pmatrix}, \quad B^\dagger=\begin{pmatrix}
        0 & 0 & \cdots &0 & 0 \\
        1 & 0 &\cdots& 0 & 0\\
        0 & 1 & \cdots &0 & 0\\
        \vdots & \vdots & \ddots& \vdots & \vdots\\
        0 & 0 & \cdots &1 & 0
    \end{pmatrix}\label{eq:hardcoreboson}
\end{equation}were introduced \cite{PhysRevA.89.012328}. They are related to the shift and clock operators by $B=X^\dagger \Pi, B^\dagger=\Pi X$, where 
\begin{equation}
    \Pi=1-\frac{1}{N}\sum_{n=1}^{N}Z^{n}=\begin{pmatrix}
        0 & 0 & 0& \cdots  & 0 \\
        0 & 1 &0 &\cdots& 0\\
        0 & 0 & 1 & \cdots & 0\\
        \vdots & \vdots & \vdots & \ddots&  \vdots\\
        0 & 0 &0&  \cdots & 1
    \end{pmatrix}  \label{eq:projection}
\end{equation}is a projection operator that satisfies $\Pi^2=\Pi$. The parafermionic statistics 
\begin{equation}
    C_jC_k=\omega C_kC_j, \quad C_j^\dagger C_k^\dagger =\omega C_k^\dagger C_j^\dagger, \quad C_jC_k^\dagger =\omega^{-1}C_k^\dagger C_j, \quad C_j^\dagger C_k=\omega^{-1}C_kC_j^\dagger, \quad \text{for }\ j<k
\end{equation}is achieved by the Fradkin-Kadanoff transformation 
\begin{equation}
    C_j= \left(\prod_{k=1}^{j-1}Z_k\right)B_j=\omega^{\sum_{k=1}^{j-1}N_k}B_j,\quad C^\dagger_j=\left(\prod_{k=1}^{j-1}Z^\dagger_k\right)B_j^\dagger=\omega^{-\sum_{k=1}^{j-1}N_k}B_j^\dagger,
\end{equation}where $B_j=\cdots\otimes1\otimes \underset{j \text{th}}{B}\otimes1\otimes \cdots$, and
\begin{equation}
    Z_k=\omega^{N_k}=1-(1-\omega)\sum_{n=1}^{N-1}\omega^{n-1}B_k^{\dagger n}B_k^n=1-(1-\omega)\sum_{n=1}^{N-1}\omega^{n-1}C_k^{\dagger n}C_k^n
    \label{eq:Zdef}
\end{equation}generalizes the identity $Z_k=1-2c^\dagger_kc_k$ for $N=2$. Since the hard-core boson operators satisfy
\begin{equation}
    B_j^N=B_j^{\dagger N}=0, \quad B_j^{\dagger n} B_j^n + B_j^{N-n} B_j^{\dagger N-n}=1, \quad \text{for }\ n=1,2,.\cdots, N-1,
\end{equation}the Fock parafermions have the desired properties of parafermionic generalization to Dirac fermions.

Fock parafermion operators are more conveniently related to the Majorana ones if we switch from the convention used in Sec.~\ref{sec:ZNparafermion} and Ref.~\cite{Fendley_2012} to that of Ref.~\cite{PhysRevA.89.012328} by changing the representation of the clock and shift operators, and define
\begin{equation}
    \Gamma_j=\left(\prod_{k=1}^{j-1}Z_k\right)X^\dagger_j,\quad \Delta_j=\Gamma_jZ_j=\left(\prod_{k=1}^{j-1}Z_k\right)X^\dagger_jZ_j.
\end{equation}They satisfy the relations
\begin{equation}
    \Gamma_j^N=\Delta_j^N=1, \quad \Gamma_j^\dagger=\Gamma_j^{N-1}, \quad\Delta_j^\dagger=\Delta_j^{N-1}, \quad  \Gamma_j\Delta_j=\omega\Delta_j\Gamma_j, \label{eq:gammadeltaparafermion1}
\end{equation}and 
\begin{equation}
   \Gamma_j\Gamma_k=\omega \Gamma_k\Gamma_j, \quad  \Delta_j\Delta_k=\omega\Delta_k\Delta_j,\quad \Gamma_j\Delta_k=\omega \Delta_k\Gamma_j,\quad \Delta_j\Gamma_k=\omega \Gamma_k\Delta_j,\quad \text{for } j<k.
   \end{equation}In therms of them, the Fock parafermions operators can be expressed as
\begin{equation}
\begin{split}
     C_j=& \Gamma_j\Pi_j= \Gamma_j-\frac{1}{N}\sum_{n=1}^{N}\omega^{\frac{n(n+1)}{2}}\Gamma_j^{n+1}\Delta_j^{\dagger n},\\
     C_j^{\dagger }=& \pi_j\Gamma_j^\dagger= \Gamma_j^\dagger-\frac{1}{N}\sum_{n=0}^{N-1}\omega^{-\frac{n(n+1)}{2}}\Delta_j^{n}\Gamma_j^{\dagger n+1}.
\end{split}   
\end{equation}The inverse transformation is given by
\begin{equation}
    \Gamma_j=C_j+C_j^{\dagger N-1},\quad  \Delta_j=C_j\omega^{N_j}+C_j^{\dagger N-1}=C_j+C_j^{\dagger N-1}+(\omega-1)\sum_{n=1}^{N-1}C_j^{\dagger n-1}C_j^n,
\end{equation}where the property $C_jC_j^{\dagger n}C_j^n=C_j^{\dagger n-1}C_j^n$ implied by \eqref{eq:anticomm} has been used. The canonical anti- and $\omega$-commutation relations of the two types of fermions and parafermions are summarized in Table \ref{tab:placeholder}.

\newcolumntype{P}[1]{>{\centering\arraybackslash}p{#1}}

\begin{table}[ht]
    \centering
    \caption{Comparison of algebraic properties of different types of fermionic and parafermionic operators.}
    \renewcommand{\arraystretch}{1.4}
    \setlength{\tabcolsep}{4pt}
    \begin{tabular}{
        P{16mm}  
        P{25mm}  
        P{21mm}  
        P{30mm}  
        P{35mm}  
    }
    \toprule
    & \textbf{Majoranas} 
    & \textbf{Fermions} 
    & \textbf{Parafermions} 
    & \textbf{Fock parafermions} \\
    \midrule

    \parbox[c]{16mm}{\centering exclusion/\\ nilpotency}
    & $a_j^2 = b_j^2 = 1$
    & $c_j^2 = c_j^{\dagger 2} = 0$
    & $\chi_j^N = \psi_j^N = 1$
    & $C_j^N = C_j^{\dagger N} = 0$ \\[2pt]
    
    \midrule

    \makecell[c]{exchange\\ statistics}
    & \makecell[c]{
        $a_j a_k = - a_k a_j$ \\
        $b_j b_k = - b_k b_j$ \\
        $a_j b_k = - b_k a_j$
      }
    & \makecell[c]{
        $c_j c_k = - c_k c_j$ \\
        $c_j^\dagger c_k^\dagger 
        = - c_k^\dagger c_j^\dagger$
      }
    & \makecell[c]{
        $\chi_j \chi_k = \omega \chi_k \chi_j$ \\
        $\psi_j \psi_k = \omega \psi_k \psi_j$ \\
        $\chi_j \psi_k = \omega \psi_k \chi_j$ \\
        $\psi_j \chi_k = \omega \chi_k \psi_j$ \\
        $\chi_j \psi_j = \omega \psi_j \chi_j$ \\
        $(j < k)$
      }
    & \makecell[c]{
        $C_j C_k = \omega C_k C_j$ \\
        $C_j^\dagger C_k^\dagger 
        = \omega C_k^\dagger C_j^\dagger$ \\
        $C_j C_k^\dagger 
        = \omega^{-1} C_k^\dagger C_j$ \\
        $C_j^\dagger C_k 
        = \omega^{-1} C_k C_j^\dagger$ \\
        $(j < k)$
      } \\[2pt]

    \midrule

    \makecell[c]{gCAR/\\ unitarity}
    & \parbox[c]{25mm}{\centering
    $a_j^\dagger a_j = a_j a_j^\dagger = 1$ \\
        $b_j^\dagger b_j = b_j b_j^\dagger = 1$
      }
    & $\{c_j, c_j^\dagger\} = 1$
    & \makecell[c]{
        $\chi_j^\dagger \chi_j 
        = \chi_j \chi_j^\dagger = 1$ \\
        $\psi_j^\dagger \psi_j 
        = \psi_j \psi_j^\dagger = 1$
      }
    &\parbox[l]{35mm}{\centering
        $C_j^{\dagger n} C_j^n 
        + C_j^{\,N-n} C_j^{\dagger\,N-n} = 1$ \\
        $n = 1, 2, \dots,N-1$
      } \\
    \bottomrule
    \end{tabular}
    \label{tab:placeholder}
\end{table}

\end{appendix}

\bibliography{DYBE.bib}
\end{document}